\documentclass{aastex63}
\usepackage[ruled,vlined]{algorithm2e}
\usepackage[utf8]{inputenc}
\usepackage{comment}
\usepackage{dsfont}
\usepackage{natbib}
\usepackage{amsmath}
\usepackage{amssymb}
\usepackage{amsfonts}
\usepackage{algorithmic}
\usepackage{graphicx}
\usepackage{textcomp}
\usepackage{booktabs}
\usepackage{bbm}
\usepackage{xcolor}
\usepackage{svg}
\usepackage{multirow}
\usepackage{dutchcal}

\accepted{\today}

\submitjournal{AJ}
\graphicspath{{./}{figures/}}
\begin{document}

\title{Efficient exoplanet imaging simulations of the Habitable Worlds Observatory}

\correspondingauthor{Jamila Taaki}
\email{xiaziyna@gmail.com}

\author[0000-0001-5475-1975]{Jamila S. Taaki}
\affiliation{Michigan Institute for Data Science, University of Michigan\\
500 S State St, Ann Arbor, MI 48109
}

\author{Farzad Kamalabadi}
\affiliation{Department of Electrical and Computer Engineering,
University of Illinois at Urbana-Champaign\\
306 N. Wright St. MC 702, Urbana, IL 61801-2918}

\author[0000-0001-6233-8347]{Athol J. Kemball}
\affiliation{Department of Astronomy, University of Illinois at Urbana-Champaign\\
1002 W. Green Street, Urbana, IL 61801-3074}

\author[0000-0002-5466-3817]{Lia Corrales}
\affiliation{Department of Astronomy, University of Michigan\
West Hall, 1085 S. University, Ann Arbor, MI 48109}

\author[0000-0002-2531-9670]{Alfred O. Hero III}
\affiliation{Department of Electrical Engineering and Computer Science, University of Michigan\
1301 Beal Avenue, Ann Arbor, MI 48109}

\begin{abstract}
Direct imaging simulations of starshades and other proposed mission concepts are needed to characterize planet detection performance and inform mission design trades. In order to assess the complementary role of a 60 m starshade for the Habitable Worlds Observatory (HWO), we develop the optical model of a starshade and simulate solar system imaging at 0° and 60° inclinations. The optical core throughput of a direct-imaging system is a key metric that governs exposure time and the potential exoplanetary yield of a mission.  We use our optical model to evaluate core throughput, incorporating 6 m segmented and obscured telescope apertures, over the visible to near-infrared wavelength band (500-1000 nm). Accurate diffractive optical simulations of this form require many large Fourier transforms, with prohibitive run-times, as both the starshade mask and telescope aperture require fine-scale spatial sampling. We introduce a Fourier sampling technique, the Bluestein Fast Fourier Transform (B-FFT), to efficiently simulate diffractive optics and enable high-fidelity simulations of core throughput. By characterizing sampling requirements and comparing B-FFT’s computational complexity to standard Fourier methods (e.g., DFT, FFT), we demonstrate its efficiency in our optical pipeline. These methods are implemented in PyStarshade \citep{Taaki2025}, an open-source Python {package} offering flexible diffraction tools and imaging simulations for starshades. Our results show the HWO starshade used with a segmented off-axis telescope aperture achieves an optimal core throughput measured within a photometric aperture of radius $0.7\lambda/D$ of 68 $\%$. With an additionally obscured aperture, a $66 \%$ core throughput is achieved.

\end{abstract}
\section{Introduction} \label{sec:intro}
The Habitable Worlds Observatory (HWO) is a future space-based mission recommended by the Astro2020 Decadal report \citep{NAP26141}, to directly image and collect broadband spectra in order to identify and characterize Earth-Sun analogs.  This task will require broadband high-contrast starlight suppression of relative order $1:10^{-8} - 10^{-10}$ \citep{traub2010direct, mennesson}. Starlight suppression systems may be separated into two categories: internal occulters (coronagraphs) and external occulters (starshades). Proposed HWO missions include coronagraph, or starshade-only missions, as well as hybrid coronagraph-starshade missions \citep{morgan2022exploration}. Optical simulations of mission designs are essential for exploring design trades in order to maximize exo-Earth yield and overall mission success. Although the starlight suppression performance of a starshade has been extensively validated in the nominal case, optical simulations of aperture-dependent performance metrics remain underexplored due to computational challenges. In this paper, we derive a general optical model of a starshade and perform simulations of optical performance metrics to assess the complementary role of a starshade to HWO. To enable these simulations, we present a fast optical propagation method and describe its computational advantages.

A starshade is a petal-shaped mask designed to fly in formation with a telescope \citep{simmons2004new, Cash_2011, cady, vanderbei2007optimal}. When aligned with a target star, the starshade produces a destructive interference pattern of on-axis starlight incident on the telescope aperture. The fractional flux suppression of the starlight is called \textit{contrast}  \citep{starshade_special_section, harness_subscale}. For every photon we receive from an Earth-like planet, we receive some $10^{10}$ photons from a Sun-like star at visible wavelengths. A starshade operates with a flux contrast of $10^{-11}$ to lower the noise floor from statistical stellar fluctuations and thereby raise the signal-to-noise ratio (SNR) of planetary photons to be detectable. While the starshade achieves excellent starlight suppression, long retarget times (days to weeks) inhibit the starshade's efficiency in characterizing a large number of potential exo-Earths. A hybrid coronagraph-starshade mission \citep{morgan2022exploration, rhonda_6m} would mitigate idle-time during retargeting and improve exo-Earth yield. In this hybrid design, a coronagraph may be used to perform a blind-search for candidate exo-Earths. Either concurrently or in a later rendezvous mission \citep{rendezvous}, the starshade follows up to obtain detailed spectral observations of these candidates. 

The \textit{optical throughput} of a starlight suppression system has been identified as a key metric for efficient retrieval of exo-Earths \citet{mennesson}. Optical throughput refers to the fraction of light from an off-axis source (i.e. a planet) that a starlight suppression system allows through. The exposure time required to image a planet for a chosen SNR is to first-order, inversely proportional to optical throughput. For an Earth-Sun analog, at a distance of 12 pc, a 60 m starshade achieves a 100 $\%$ throughput. \textit{Core throughput} refers to the intensity in the central core of the point-spread-function of a planet-like source. A starshade suppresses starlight before it enters the telescope aperture. Therefore, the optical throughput achieved by the starshade is decoupled from the features of the telescope aperture \citep{mennesson}. However, the point-spread-function results from propagating the diffracted light past the telescope aperture, therefore, core throughput will be affected by the telescope aperture. Core throughput influences exposure times, predicted yield, and therefore mission design. For coronagraphs, obscurations and, segmentation are known to strongly affect the expected exo-Earth yield \citep{stark_2019}. Despite it being a significant optical metric \citep{hwo_ss}, the core throughput of a starshade has not yet been characterized for the proposed HWO segmented, or obscured, telescope aperture as may be used in a hybrid-mission, and to date has only been calculated for an ideal circular aperture. Hence, there is a need for numerically efficient techniques to perform optical simulations with HWO apertures for a segmented and on- or off-axis aperture.

End-to-end optical modeling of a starshade is numerically intensive, as the point-spread-function varies both spatially and spectrally. Because it is numerically intensive, a telescope aperture is typically modeled with a pixel scale of several cm, despite a fine-scale sampling of the starshade mask itself. For example in the Starshade Imaging Simulation Toolkit for Exoplanet Reconnaissance (\textit{SISTER}) the Roman telescope aperture is modeled with $\sim 5$ cm scale pixel sizes \citep{sisters} and with a 16 cm pixel-size in \citet{barnett2020efficient}. The HWO segmented aperture designs have cm scale features and thereby require sampling at this precision. These fine-scale features result in broadened point-spread-functions and a potential loss of core throughput, changing the imaging exposure time requirements. Our work focuses on the application of the Bluestein Fast Fourier Transform (B-FFT hereinafter) \citep{bailey1991fractional} to efficiently calculate spatial and spectrally dependent PSFs with a fine spatial sampling of the telescope apertures.

Starshade imaging may be modeled using classical Fourier optics where complex-valued optical fields are propagated between optical planes using Fourier integral representations of optical diffraction \citep{goodman2005introduction, Born_Wolf}. A general model for calculating the spatially varying PSF of an astrophysical source requires two stages of optical propagation. The first stage of the optical pipeline calculates the diffraction at the entrance of the telescope. The second stage propagates the incident light past the aperture lens and into the focal plane. These calculations are performed over a range of wavelengths and source positions to generate a basis collection of PSFs to simulate imaging for astrophysical scenes \citep{sisters}. 

Starshade diffraction incident on the telescope is in the near-field Fresnel regime and may be described by the convolution of the mask with a Fresnel kernel \citep{goodman2005introduction}. Broadly, to accurately simulate flux contrast levels of $10^{-11}$ with equispaced sampling requires a starshade pixel sampling $\Delta s$ that is sufficiently small (mm-scale) to control aliasing error below the $10^{-11}$ contrast level \citep{barnett2020efficient, aime_ss}. This creates an inherent tradeoff between choices of imaging parameter samplings or resolutions and run-time scaling. There are several approaches to efficiently calculate starshade diffraction at the entrance of the telescope aperture that utilize non-equispaced quadrature methods \citep{Papalex}, including the boundary diffraction wave (BDW) method \citep{cady2012boundary, harness2018advances}. In the BDW method, Green's theorem is applied to obtain the diffraction at a position in the aperture plane from a one-dimensional summation-integral over the edge points of the occulting starshade mask, this approach is also known as a Maggi–Rubinowicz representation \citep{Born_Wolf}. \citet{barnett2020efficient} derive a non-uniform sampling scheme for the starshade mask, which, when combined with the Non-Uniform FFT (NU-FFT), achieves the required contrast precision using fewer samples than a uniform grid. The non-uniform sampling is denser near the starshade edges—where the Fresnel kernel is more oscillatory—and coarser near the center. \citet{aime_ss} utilize two-step Fresnel filtering, which admits a lower-sample complexity computed in the Fourier spectral space owing to the low-frequency content of the starshade diffraction. These starshade diffraction methods are efficient for calculating the field at the aperture for a binary starshade mask. While efficient non-uniform sampling schemes can be derived for analytic masks under plane wave illumination, they are not applicable for propagating arbitrary fields as arise in multi-stage optical simulations. An efficient algorithm for non-uniform sampling at the aperture is needed in order to calculate the PSF. However, this requires full knowledge of the incident optical field. SISTER \citep{sisters} is implemented in Matlab and uses the BDW method for starshade diffraction; however, it applies the Discrete Fourier Transform (DFT) \citep{golub1983matrix} to propagate the aperture field to the focal plane to generate a PSF basis for imaging simulations. 

Simulating imaging, as well as calculating core throughput, requires resolving Fourier-transform calculations across different windows and spatial scales on finely sampled grids. Using a direct FFT or DFT to propagate the high-resolution telescope field to the focal plane is computationally expensive. Our work applies the B-FFT (also called the Chirp Z transform) to efficiently move between resolutions and windows in Fourier-optical propagation, first shown to benefit optical simulations in \citet{leutenegger2006fast} and described in \citet{jurling2018techniques, hu2020efficient}. The B-FFT overcomes the FFT's fixed spatial-spectral sampling constraint, but benefits from the numerical efficiency of an FFT. \citet{Soummer07} identify the potential utility of a B-FFT (referred to as the fractional FFT) to perform direct-imaging simulations for Lyot-style coronagraphs. Concurrent software, \textit{HCIPy} \citep{hcipy}, supports this method (called the zoom-FFT) in simulating coronagraphs. 

Despite the numerical efficiency of the B-FFT, it has not been used for starshade optical simulations. Here, we justify its use in the context of HWO starshade simulations with a theoretical and empirical analysis of sampling requirements and computational efficiency. Our optical pipeline involves two stages of optical propagation: first propagating diffracted light past the starshade to the telescope aperture (large windows to small window), and second calculating a PSF basis for a grid of source positions (same sized input-output). We apply the B-FFT to both stages, where zero-padding by a factor of $\sim 10 \times$ or more rules out using a direct FFT. Both stages of the pipeline are of comparable computational requirements. In the context of studying apertures, the first stage is sufficient for simulation purposes and generalizable to arbitrary masks and the second stage has a $6 \times$ reduction in floating-point operations compared to the DFT. 

These diffraction tools are implemented in an open-source Python {package}, \textit{PyStarshade} \citep{Taaki2025}, as well as tools to perform optical simulations of an exoplanetary scene imaged with a starshade. The simulation framework introduced here has two complementary use cases: (i) nominal instrument performance trade and yield studies across starshade designs and apertures and (ii) analyzing the tolerance to perturbations or defects in the starshade or the telescope. PyStarshade uses starshade mask design files from SISTER \citep{sisters} and interfaces with HCIPy \citep{hcipy} to download and utilize various telescope apertures provided by the library. PyStarshade also contains examples to utilize exoplanetary scenes simulated with \textit{ExoVista} \citep{Stark_2022} for imaging. 

We perform HWO simulations with a 60m starshade concept \citep{hwo_ss} operating in the visible-NIR (500-1000 nm) with a 6 m telescope aperture. This band allows for spectral characterization of exo-Earth gases (CO2, H2O, CH4, O2, and H2) \citep{Latouf_2023}. We demonstrate that the starshade achieves the optimal core throughput for an off-axis aperture equal to $68 \%$ and close to optimal for an on-axis {obscured} pupil equal to $66 \%$ as compared to an ideal circular aperture with a theoretical core throughput of $68 \%$, within a photometric aperture of radius $0.7 \cdot \lambda / D$ where $\lambda / D$ is the angular resolution for wavelength $\lambda$ and effective telescope diameter $D$ for either aperture. Since the exposure time to achieve a specified signal-to-noise ratio is approximately inversely proportional to throughput, this $\sim 2 \%$ relative difference in throughput implies that the starshade with an obscured pupil requires an increase in exposure time of approximately $3 \%$. Consequently, our analysis establishes that the choice of aperture does not strongly impact exposure time and that starshades maintain their expected exo-yield with these non-circular aperture designs. To demonstrate the imaging capabilities of a starshade augmented HWO telescope, we perform imaging simulations of our solar system using scenes simulated with ExoVista \citep{Stark_2022} at two different inclinations. At 0 degrees (face-on) and at 60 degrees.

This paper is organized as follows. Section \ref{sec: methods} describes the starshade optical model and computation method. In Section \ref{sec: results} we report the results from our simulations of the HWO starshade concept. Section \ref{sec: discussion} contains a discussion of core throughput performance and conclusions are presented in Section \ref{sec: conclusion}.

\section{METHODS} \label{sec: methods}
This section describes a general optical model of a starshade and optical sampling requirements. In the described Fourier optics model, boldface symbols, such as $\mathbf{x}$, denote column vectors, e.g., representing spatial coordinates $(x, y)$. When describing the optical propagation at each stage, $(\zeta, \eta)$ are the input spatial coordinates, and $(x, y)$ are the output spatial coordinates.

The calligraphic $\mathcal{F}$ denotes the continuous 2D Fourier transform defined for an input function $f(\mathbf{x})$, over spatial coordinates $\mathbf{x} = (x, y)^T$, onto frequency coordinates $\boldsymbol{\omega} = (\omega_x, \omega_y)^T$, as $\mathcal{F}(f(\mathbf{x}))(\boldsymbol{\omega}) = \iint_{-\infty}^{\infty} f(\bold{x})  e^{-j 2 \pi {\boldsymbol{\omega}^T \bold{x}}}  \,d\bold{x}$. The inverse 2D Fourier transform is  $\mathcal{F}^{-1}\left(f(\boldsymbol{\omega}) \right)(\mathbf{x}) = \iint_{-\infty}^{\infty} e^{j 2\pi \boldsymbol{\omega}^T \mathbf{x}}f(\boldsymbol{\omega}) \,d\boldsymbol{\omega}$ \citep{blahut2004theory}. We use $\mathbb{E}$ to denote expectation. Notation is provided in Table~\ref{tab:symbols_ss}. 

\begin{table}[htbp!]
\footnotesize
\caption{List of symbols}
\begin{center}
\begin{tabular}{r l p{10cm} }
\toprule
$s(x, y)$ & Starshade mask\\
$R_{ss}$ & Starshade radius\\
$P(x, y)$ & Telescope aperture mask\\
$D_{P}$ & Telescope aperture diameter \\
$\lambda$ & Wavelength \\
$d_0$ & Farfield distance from exoplanetary system  to starshade \\
$d$ & Distance between starshade and telescope \\
$d_f$ & Focal length of telescope \\
$IWA = \frac{R_{ss}}{z}$ & Inner working angle \\
$\theta_{P} = \frac{\lambda}{D_P}$ & Telescope angular resolution \\
$\mathcal{f} = \frac{R_{ss}^2}{\lambda z}$ & Starshade Fresnel number  \\
$\Delta x, N_x$ & Source field pixel size, number of samples along one axis \\
$\Delta s, N_s$ & Starshade mask pixel size, number of samples along one axis of starshade \\
$\Delta P, N_P$ & Telescope aperture pixel size, number of samples along one axis of pupil\\
$\Delta f, N_f$  & Output image pixel size, numer of imaging pixels\\
\bottomrule
\end{tabular}
\end{center}
\label{tab:symbols_ss}
\end{table}

\subsection{Optical model} \label{sec:fresnel} \label{sec:optics}
\begin{figure}[h!]
  \centering
  \includegraphics[width=\linewidth]{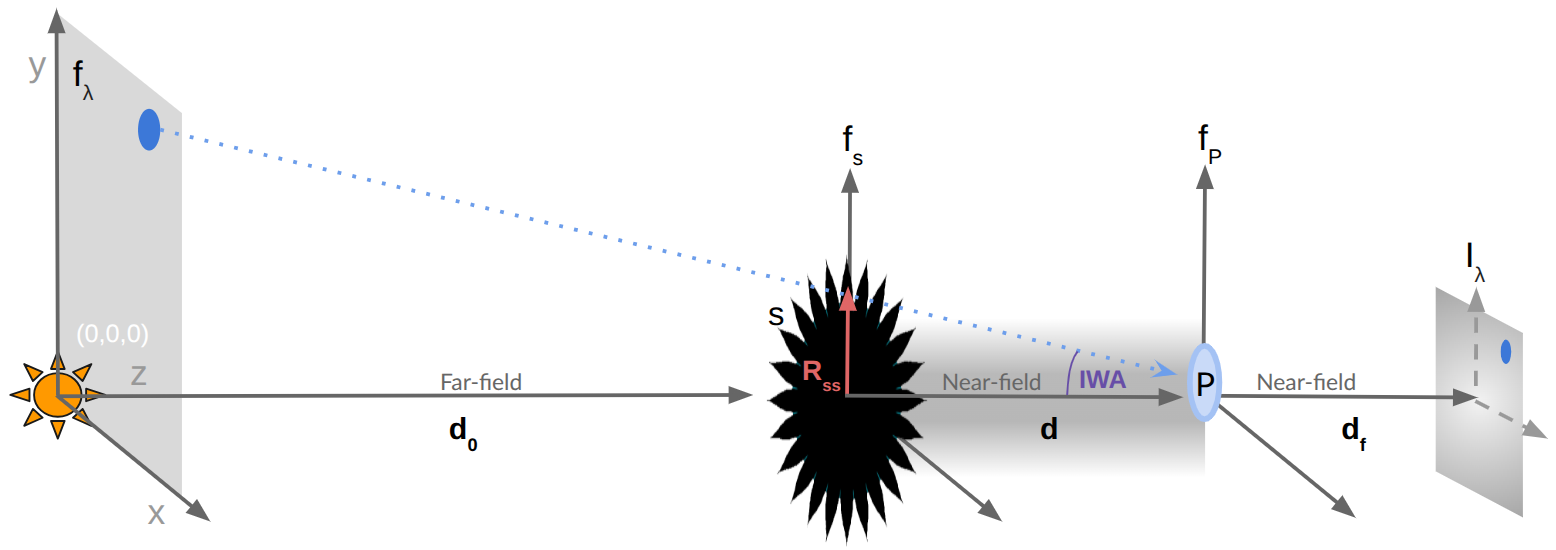}
    \caption{The optical model of a starshade is illustrated. The input function $f_{\lambda}(x, y)$ describes the flux density of a star-planet scene, as viewed from a fixed angle, and as a function of wavelength. This input function propagates past the starshade $s(x,y)$, onto a telescope aperture mask $P(x, y)$ at distance $d$, and onto a CCD in the image plane at a distance $d_f$. The IWA is the angular separation from the star where an exoplanet can be imaged. In the output image $I_\lambda(x,y)$, the light from the host-star has been suppressed revealing the faint exoplanetary system.} \label{fig: starshade_im}
\end{figure}

We adopt a general Fourier optics model to describe the propagation of light from an exoplanetary source scene through the starshade optics, to produce the imaged scene $I_\lambda (\mathbf{x})$ (Figure \ref{fig: starshade_im}). This general model is implemented in varying formulations; \citet{mengya_hu, sisters, aime_ss}. The optical model consists of a sequence of plane parallel fields defined over wavelength $\lambda$ along the optical axis $z$ with the center at $(x, y) = 0$. The source field lies at the origin $z = 0$. 

First, light from the source scene is modeled as a collection of statistically independent sources, each propagates independently through the starshade optical model. In Figure \ref{fig: starshade_im}, the input source field $f_{\lambda} (\bold{x})$ describes the flux of an exoplanetary scene at wavelength $\lambda$. The source field is discretized into coordinates $\mathbf{x}_k : k \in K$, a source pixel is denoted as $f_{\lambda, k} := f_{\lambda}(\bold{x}_k)$ representing the flux within a unit pixel area and behaves as a point source.  If a source is at a position $\mathbf{x}_k$, the angular source position is $\boldsymbol{\phi}_k = \frac{\mathbf{x}_k}{d_0}$. The independence of spatially separated sources implies they are statistically uncorrelated: $\mathbb{E}[f_{\lambda, k}^* f_{\lambda, j}] = \delta(k - j) |f_{\lambda, k}|^2$ \citep{blahut2004theory}. The propagated intensity of a point source defines a point-spread-function (PSF) in the focal plane and is denoted as $\psi_{\lambda, k}(\bold{x})$. 

The imaged scene $I_\lambda (\mathbf{x})$ at the focal plane at a single wavelength $\lambda$ is found as the incoherent sum of PSFs weighted by source flux pixels \citep{sisters}:
\begin{align} \label{eq: psf_sum}
\mathbb{E}[I_\lambda](\mathbf{x}) = \sum_k f_{\lambda, k} \psi_{\lambda, k}(\mathbf{x} - \boldsymbol{\phi}_k d_f).\end{align}
Where $d_f$ is the focal length of the telescope and controls the magnification of the imaged scene. Similarly, a band-pass image is obtained by incoherently summing the images $\{ I_\lambda : \lambda \in B\}$ over the band $B$.

We now describe the optical derivation of the spatial-spectral dependent PSF basis $\{ \psi_{\lambda, k}(\mathbf{x}): k \in K\}$, as the key component needed to simulate imaging in Equation \ref{eq: psf_sum}. The light from a source undergoes three stages of optical propagation to produce the PSF intensity. In the first step of the optical model, light from a source propagates through free space and is incident on a starshade. A starshade is of radius $R_{ss} \sim 10-100$ m, while the distance $d_0$ between a typical exoplanet scene and starshade is of the order of parsecs. Therefore, the exoplanet scene is in the optical far-field  ($d_0 \gg \frac{R_{ss}^2}{\lambda} $). A point source situated in the exoplanetary scene propagates over $d_0$ and becomes a plane wave incident on the starshade. The field incident on the starshade is denoted here by $f_{s, \lambda}(\mathbf{x})$. The field immediately after masking by the starshade $s(\mathbf{x})$ is:
\begin{align} \label{eq: 2}
f_{s', \lambda}(\mathbf{x}) = s(\mathbf{x}) \cdot f_{s, \lambda}(\mathbf{x}).
\end{align}
The masked plane wave undergoes Fresnel diffraction over a near-field distance $d  \sim 10^7$ m to the telescope aperture. Starshades are designed to operate over moderate Fresnel numbers $ \mathcal{f} = \frac{R_{ss}^2}{\lambda d} \sim 10 - 20$. 
Fresnel diffraction may be expressed in various equivalent Fourier forms, with different computational approaches  \citet{goodman2005introduction}. See \cite{aime_ss} for an approach to calculating starshade diffraction using a forwards-backwards Fourier transform. The form used here and by \citet{cady2012boundary}, \citet{harness2018advances}, and \citet{barnett2020efficient} expresses Fresnel diffraction with a single Fourier-transform. Using this form, the incident field at the aperture is denoted as $f_{P, \lambda}(\mathbf{x})$ and is given by:
\begin{align} \label{eq: fresnel_telescope}
f_{P, \lambda}(\mathbf{x}) = \frac{e^{ \frac{j 2 \pi d}{\lambda}}}{j \lambda d} e^{\frac{j \pi \mathbf{x}^T \mathbf{x}}{\lambda d}} \mathcal{F} \left( f_{s', \lambda}(\boldsymbol{\zeta}) e^{\frac{j \pi \boldsymbol{\zeta}^T\boldsymbol{\zeta}}{\lambda d} } \right) \left( \frac{\mathbf{x}}{\lambda d} \right).
\end{align} The domain of the Fourier integral is over the starshade mask and depends on $\lambda d$, through the Fresnel diffraction kernel.

\begin{figure}[h!]
  \centering
  \includegraphics[width=.5\linewidth]{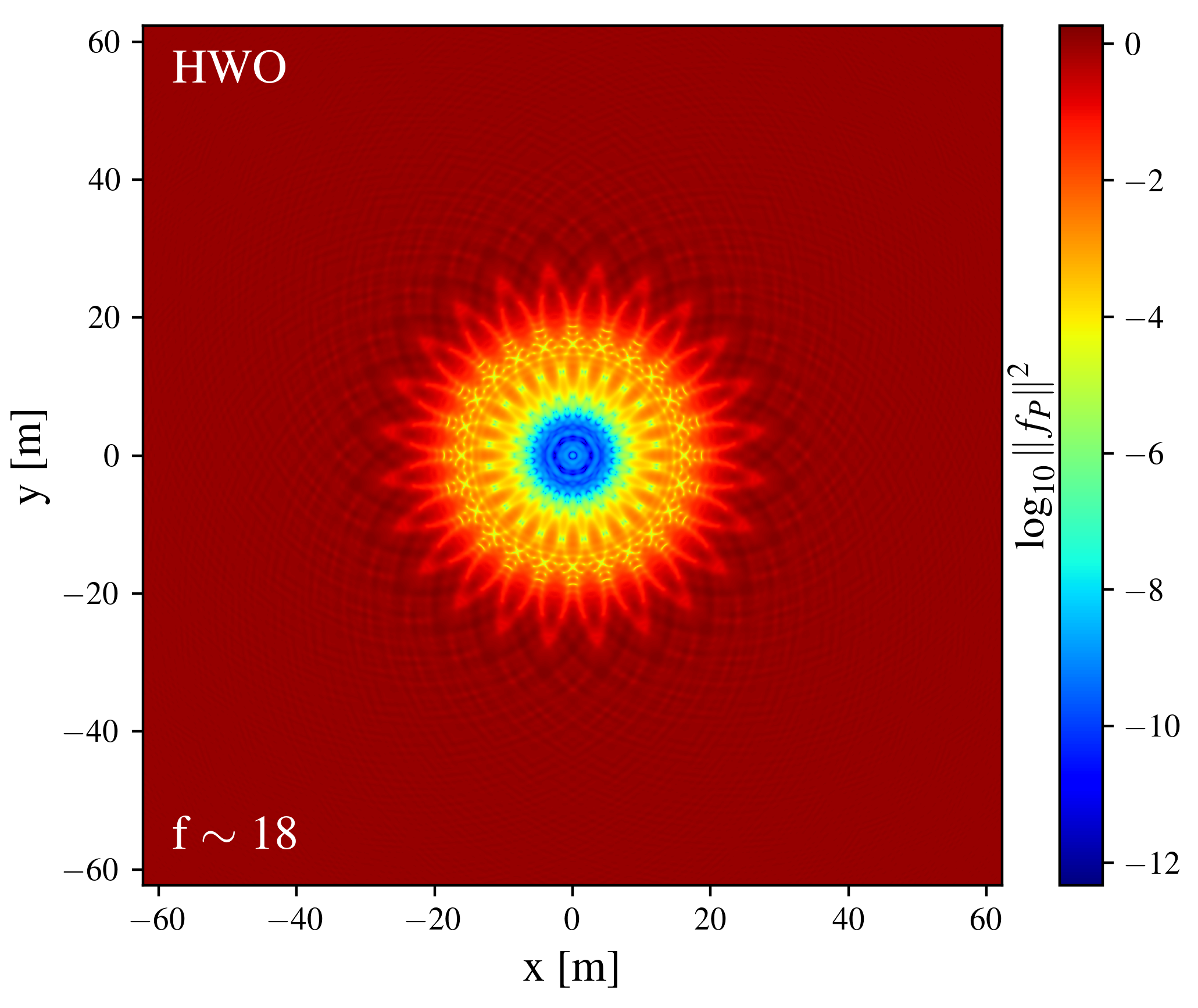}
    \caption{Suppression in logarithmic units (diffraction intensity at the telescope aperture $\|f_P^{\parallel}\|^2$) is shown for the HWO 60 m starshade with 16 petals at a wavelength of 500 nm, with a spatial sampling of $\Delta P=2$ cm, $N_P = 6000$. The central dark shadow is masked by the telescope and produces the on-axis stellar PSF. The dark shadow must therefore be as large as the aperture. To maintain the dark shadow (contrast), the Fresnel number should remain constant $\mathcal{f} = \frac{R_{ss}^2}{\lambda z} \sim 10-20$. This can be maintained at wavelengths outside of the band by changing the lateral flight distance of the starshade $z$. The dark shadow is approximately the size of the starshade in diameter (60m). } \label{fig: darkshadow}
\end{figure}

An example of a diffracted field incident on a telescope aperture $f_{P, \lambda}$ is shown in Figure \ref{fig: darkshadow}. The field is suppressed in a region around the axis referred to as the \textit{dark shadow} \citep{starshade_special_section}. Semi-analytic approximations of $f_{P, \lambda}$ can be obtained for an analytic form of transmission function $s(\mathbf{x})$; see \citet{Vanderbei_2003, vanderbei2007optimal, Cash_2011} for a Bessel-function expansion. For a non-analytic, petal-shaped mask, a numerical diffraction calculation is necessary.

The telescope is modeled as a thin lens with a focal length $d_f$.
If the telescope is treated as a thin lens, the field immediately after the telescope $f_{P', \lambda}(\mathbf{x})$ is the product of the incident field $f_{P, \lambda}(\mathbf{x})$ masked by a telescope aperture mask $P (\bold{x})$ and a phase function $\theta_{\lambda} (\mathbf{x}) = e^{\frac{- i \pi \mathbf{x}^T \mathbf{x}} {\lambda d_f}}$ for telescope of focal-length $d_f$ \citep{goodman2005introduction}. Because of this quadratic phase term, Fresnel propagation from the aperture to the focal plane simplifies to Fraunhofer diffraction of the masked field $f_{P', \lambda}$ without the phase term \citep{blahut2004theory, goodman2005introduction}: 
\begin{align} \label{eq: focal}
f_{f, \lambda}(\mathbf{x}) = \frac{e^{\frac{j 2\pi d_f}{\lambda}}}{j \lambda d_f} e^{\frac{j \pi \mathbf{x}^T \mathbf{x}}{\lambda d_f}} \mathcal{F} \left( f_{P, \lambda}(\boldsymbol{\zeta}) \cdot P(\boldsymbol{\zeta}) \right) \left( \frac{\mathbf{x}}{\lambda d_f} \right). 
\end{align}
The PSF $\psi_\lambda(\mathbf{x})$ formed at the focal plane at wavelength $\lambda$ is the intensity of the field at the focal plane $f_{f, \lambda}(\mathbf{x})$:
\begin{align} \label{eq: psf}
\psi_{\lambda}(\bold{x}) = |f_{f, \lambda}|^2 (\mathbf{x}).
\end{align}
A unit amplitude on-axis plane wave incident on the starshade has no phase gradient and the field past the starshade is of the form $f_{s', \lambda}(\mathbf{x}) \propto  s(\mathbf{x})$. The starshade diffraction for the on-axis source in Equation \ref{eq: fresnel_telescope} is then directly the Fresnel diffraction of the starshade mask:
\begin{align} \label{eq: fresnel_telescope_2}
f^\parallel_{P, \lambda}(\mathbf{x}) = \frac{e^{\frac{j \pi \mathbf{x}^T \mathbf{x}}{\lambda d}}}{j\lambda d} \mathcal{F} \left( s(\boldsymbol{\zeta}) e^{\frac{j \pi \boldsymbol{\zeta}^T \boldsymbol{\zeta}}{\lambda d}} \right) \left( \frac{\mathbf{x}}{\lambda d} \right).
\end{align}
We drop the constant phase terms in Equation \ref{eq: fresnel_telescope} as these bear no effect on the imaged intensity.
In order to numerically compute propagation, optical fields must span finite domains. However, the starshade mask $s(\mathbf{x})$ in Equation \ref{eq: fresnel_telescope_2} has infinite spatial extent since light is blocked within the mask and passed outside. Windowing this field leads to spectral leakage\footnote{Windowing is equivalent to masking our field with a square mask $\Pi$ i.e. $f_{s,\Pi} = f_s \cdot \Pi$. The Fourier transform of the windowed field is  $\mathcal{F}(f_{s,\Pi}) = \mathcal{F}(f_s) * \mathcal{F}(\Pi)$ where $*$ is the convolution operator. This has the effect of convolving the original Fourier transform of the field $\mathcal{F}(f_s)$ with $\mathcal{F}(\Pi)$, a squared sinc function. This smears $\mathcal{F}(f_s)$ and reduces the frequency resolution.}. Starshade optical propagation approaches \citep{Cash_2011, cady2012boundary, hu2017simulation, harness2018advances, barnett2020efficient, aime_ss} use Babinet's principle described in the monograph by \citet{Born_Wolf}. Babinet's principle applied here means that the diffracted starshade field is equivalent to the field diffracted past the complement of the starshade mask  $s^\perp (\mathbf{x})= 1 - s(\mathbf{x})$ subtracted from the field propagated through free space, modifying Equation \ref{eq: fresnel_telescope_2} to:

\begin{align} \label{eq: fresnel_telescope_3}
f^\parallel_{P, \lambda}(\mathbf{x}) = \frac{1}{\lambda d} \left(1 - e^{\frac{j \pi \mathbf{x}^T \mathbf{x}}{\lambda d}}\mathcal{F} \left( s^\perp(\boldsymbol{\zeta}) e^{\frac{j \pi \boldsymbol{\zeta}^T \boldsymbol{\zeta}}{\lambda d} } \right) \left( \frac{\mathbf{x}}{\lambda d}\right)  \right).
\end{align}

The point-spread-function of an off-axis point source can be computed from the diffracted intensity at the telescope for an on-axis source \citep{aime_ss}. Omitting amplitude, an off-axis source at an angular position relative to the observer $\boldsymbol{\phi} = (\phi_x, \phi_y)^T$ produces a plane wave incident on the starshade of the form: 
\begin{align} \label{eq: field_ss}
f_{s, \lambda} (\bold{x})  \propto e^{\frac{-j 2 \pi \mathbf{x}^T \boldsymbol{\phi}}{\lambda}}. 
\end{align}
Applying the Fourier phase-shift relation  \citep{blahut2004theory} to Equation \ref{eq: fresnel_telescope}, an off-axis plane wave produces a shift of the intensity $f^\parallel_{P, \lambda}$ at $\frac{\boldsymbol{\phi} \lambda d}{\lambda} = \boldsymbol{\phi} d$, post multiplied by a modulating term of the form $e^{\frac{j 2 \pi \mathbf{x}^T \boldsymbol{\phi}}{\lambda}}$. The field incident on the telescope aperture is of the form:
\begin{align} \label{eq: shift_psf}
f^{\boldsymbol{\phi}}_{P, \lambda} (\mathbf{x}) = e^{\frac{j 2 \pi \mathbf{x}^T \boldsymbol{\phi}}{\lambda}} f^\parallel_{P, \lambda}\left( \mathbf{x} - \boldsymbol{\phi} d \right).
\end{align}

Therefore, to compute a PSF basis $\{ \psi_{\lambda, k}(\mathbf{x}) : k \in K \}$, the starshade diffraction at the telescope aperture need only be calculated once per wavelength for an on-axis source $f^\parallel_{P, \lambda}(\mathbf{x})$, with regions of this field independently propagated past the telescope aperture to the focal plane for different off-axis positions $\boldsymbol{\phi}$. By dropping the modulating term $e^{\frac{j 2 \pi \mathbf{x}^T \boldsymbol{\phi}}{\lambda}}$ in Equation \ref{eq: shift_psf}, in Equation \ref{eq: focal} and Equation \ref{eq: psf} we obtain the PSF for an off-axis source centered at $(x, y) = (0, 0)$ in the focal plane:

\begin{align}
\psi_{\lambda, k}(\mathbf{x}) = \frac{1}{\lambda^2 d_f^2} \left| \mathcal{F} \left( f^\parallel_{P, \lambda}\left( \boldsymbol{\zeta} - \boldsymbol{\phi}_{k} d \right) \cdot P(\boldsymbol{\zeta})\right)\left(\frac{\mathbf{x}}{\lambda d_f} \right) \right|^2.
\end{align}

Figure \ref{fig: darkshadow} is an example of the field at the telescope aperture $f_P^\parallel(\mathbf{x})$ showing a central dark-shadow for HWO. Since the field at the aperture for an off-axis source is found from shifting the field for an on-axis source, $f_P^\parallel(\mathbf{x})$, light is suppressed within a limited region of a source field, less than twice the inner-working-angle (IWA) (IWA $\approx \frac{R_{ss}}{d}$) \citep{starshade_special_section, sisters}.  

\subsection{Sampling and Efficient Diffraction} \label{sec: numerical} \label{sec: pupil_sampling}

In the general optical model above, diffraction is described using continuous Fourier transforms. However, to perform numerical simulations, these expressions must be discretely sampled as Discrete Fourier Transforms (DFTs). The chosen spatial sampling of an input field determines the accuracy of the diffracted field through the Shannon-Nyquist sampling theorem \citep{oppenheim1997signals}. If a field is undersampled, regions of the diffracted field will fold onto other regions and corrupt the simulated diffraction pattern, seen as degraded contrast at the telescope aperture and spurious intensity artifacts in the PSF. When there are sampling requirements at multiple stages of optical propagation, using a DFT or FFT to meet them can be computationally costly, with run-time scaling poorly with size. For instance, achieving an output sampling with an FFT often requires zero-padding the input, which increases both memory usage and computational complexity. Here, we describe numerical sampling requirements to accurately perform these optical computations and motivate a more efficient approach using the B-FFT. A comparison of the computational complexity between the B-FFT and alternate Fourier methods is provided in Table \ref{tab:comparison}.

The optical simulation pipeline is designed to be modular and consists of three sequential stages. First, the diffracted field at the telescope aperture $f_{P, \lambda}^{\parallel}(\mathbf{x})$ given in Equation \ref{eq: fresnel_telescope_2} is computed for a specified starshade. Second, a PSF basis is computed for point sources at different off-axis positions for a particular choice of telescope aperture $P(\mathbf{x})$. Third, optical simulations of a discretized exoplanetary scene are performed. 

As described in Section \ref{sec:optics} the imaging basis consists of a collection of PSF's simulated for every off-axis source position. We compute this over a source field region extending $4 \cdot IWA$ along either axis and centered on the star \citep{iwa_cady}.  To compute a PSF basis as described in Section \ref{sec:optics}, spatial segments of the field at the telescope aperture of an on-axis plane wave $f^\parallel_{P, \lambda}$ centered on $\boldsymbol{\phi}\cdot d$ are each masked by the telescope aperture $P(\mathbf{x})$ and propagated to the focal plane. To propagate a scene at a particular wavelength (Equation \ref{eq: psf_sum}) we propagate the field for source pixels with the region of starlight suppression by scaling each PSF $\psi_{\lambda, k}(\mathbf{x})$ by the source pixel flux $f_{\lambda, k}$ and co-adding the PSFs. Outside of this radius we propagate the source pixels by convolving them with a single off-axis PSF. 

We now describe the sampling requirements for the optical propagation calculations driven by the small-scale features of the starshade and telescope aperture. The starshade mask sample size is $\Delta s$ with $N_s$ samples across the mask along one axis and the telescope aperture sample size is $\Delta P$ with $N_p$ samples along the aperture. The final image sample size is $\Delta f$ with $N_f$ pixels. For simplicity we assume the same sampling along vertical and horizontal directions. 

The field at the aperture $f^\parallel_{P, \lambda}$ is of size $N^\parallel_P = \frac{4 \cdot IWA \cdot d }{\Delta P} = \frac{4 \cdot R_{ss}}{\Delta P}$. We choose the source field sampling, and therefore PSF basis size, to generally be smaller than the telescope angular resolution. In the following, we generate a PSF basis with a source pixel size of $\Delta \phi = 2$ mas and a focal-plane pixel size of $\Delta f =$  $2$ mas. For an aperture diameter $D_P = 6$ m telescope, this sampling corresponds to $10-20$ pixels per PSF resolution element $\frac{\lambda}{D_P}$. Each PSF term $\psi_{\lambda, k}(\mathbf{x})$ is highly concentrated spatially, but does not fall off to zero so the spatial extent of $\psi_{\lambda, k}$ is chosen to empirically achieve a certain level of accuracy. Here, the PSF for a source at $(x,y) = (0,0)$ is simulated over a region of $160 \frac{\lambda}{D_P}$ to capture residual stellar brightness and $> 99.9 \%$ of the PSF intensity. However, off-axis sources are sampled over a region of $20 \frac{\lambda}{D_P}$, corresponding to $> 99 \%$ of the PSF intensity.

Equation \ref{eq: fresnel_telescope_2} describes the starshade Fresnel diffraction; the starshade mask $s(\mathbf{x})$ is multiplied by an oscillatory chirp term $e^{\frac{j \pi (\eta^2 + \zeta^2)}{\lambda d}}$ and the diffracted field is obtained from the Fourier transform of the masked chirp. The oscillations of this chirp term increase in frequency with radius, with the highest frequency oscillations of the order of $O(\frac{1}{\mathcal{f}})$ where the Fresnel number is $\mathcal{f} = \frac{R_{ss}^2}{\lambda d} \sim 10 - 20$ \citep{barnett2020efficient}. As described in \citet{Papalex} when diffraction is calculated over a uniformly sampled grid a very small sampling of our starshade mask, $\Delta s$, is required to sample the chirp and therefore minimize aliasing error to accurately compute the diffracted field at the aperture. Accordingly, the ideal propagation approach depends on the Fresnel number \citep{Papalex}. The starshade diffraction spectrum (Fourier transform) is non-bandlimited, thus some aliasing error will always occur \citep{oppenheim1997signals}. The HWO $D_{ss} = 60$ m starshade studied in this work is described in Section \ref{sec: starshade_mask}. In Section \ref{sec: starshade_mask}, a starshade spatial sampling $\Delta s = 1 $ mm is found empirically to maintain diffraction error below the $10^{-10}$ contrast level. 

In general, the sampling of the telescope aperture $\Delta P$ must also be chosen to minimize aliasing error in the PSF basis \citep{sisters}. Applying the Shannon-Nyquist sampling criterion, the maximum sampled spatial position of the PSF as determined by $\Delta P$ should encircle a significant fraction of the intensity, thus $\frac{1}{\omega_{|\psi_{\lambda, k}|^2<99.9 \%}} < \Delta P$. In this work, the cm-scale features of the studied apertures induce a more stringent sampling constraint than for a circular aperture and require a smaller $\Delta P$. For the telescope aperture sampling $\Delta P$, the spectral output samples of the Fourier transform in Equation \ref{eq: fresnel_telescope_2} are evaluated at $\Delta \omega = \frac{\Delta P}{\lambda d}$. 

\begin{figure}[h!]
  \centering
  \includegraphics[width=0.9\columnwidth]{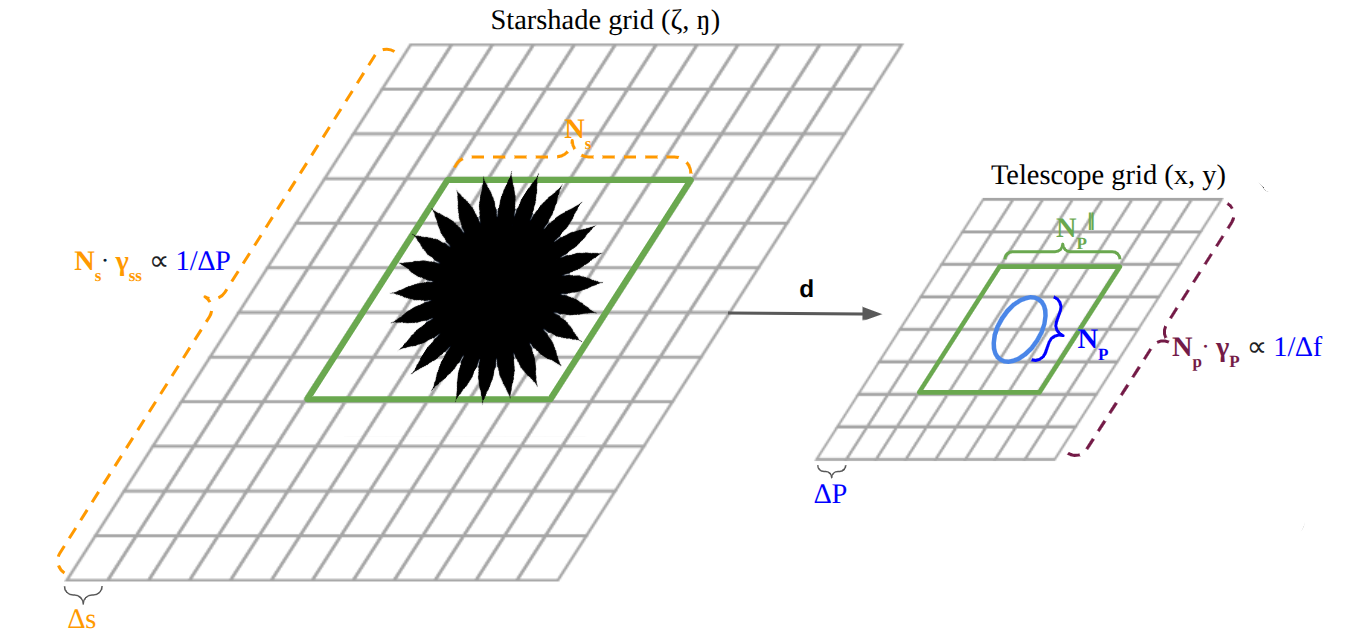}
    \caption{Illustration of the starshade and pupil plane sampling (not to scale). The pupil radius is approximately $\frac{1}{10}$ the starshade radius. The starshade grid is of size $N_s \cdot N_s$ with sampling $\Delta s$. The zero-padding of the starshade $N_s \cdot \gamma_{ss}$ necessary to achieve a $\Delta P$ sampling in the telescope plane with an FFT computation is illustrated. In the telescope plane, the telescope aperture of size $N_P \cdot N_P$ is shown, as is the slightly larger computed field of size $N_P^\parallel \cdot N_P^\parallel$.} \label{fig: sampled_field}
\end{figure} 

If Equation \ref{eq: fresnel_telescope_2} is computed with an FFT, there are a fixed number of input and output samples $N$ with output sampling $\Delta \omega = \frac{\omega_{max}}{N}$. If the starshade mask is discretely sampled at $\Delta s$, the maximum frequency of a Fourier transform of $s(\mathbf{x})$ is $\omega_{max} = \frac{1}{\Delta s}$. The choice of $\Delta s$ is driven by the contrast error constraint. The FFT with uniform spatial samples spaced by $\Delta P$ requires a zero-padded FFT of $s$ to size $N$:
\begin{align}
N = \frac{\omega_{max}}{\Delta \omega} = \frac{ \lambda  d}{ \Delta s \Delta P}.
\end{align} \label{eq: sample_pupil}
Defining a zero-padding factor $\gamma_{ss}$, a zero-padded size along one axis is $N = \gamma_{ss} N_s$. Since $N_s \cdot \Delta s = D_{ss}$, where $D_{ss}$ is the starshade diameter, we have:
\begin{align} \label{eq: zp_ss}
\gamma_{ss} = \frac{\lambda d}{ \Delta P \cdot D_{ss}}.
\end{align}

Similarly a zero-padding factor $\gamma_P$ for propagating from the telescope aperture to the focal plane: 
\begin{align} 
\gamma_P = \frac{\lambda d_f}{ \Delta f \cdot D_P},
\end{align}

Figure \ref{fig: sampled_field} illustrates the starshade and pupil plane sampling. The starshade zero-padding factor $\gamma_{ss}$ in Equation \ref{eq: zp_ss} ranges here between 40$-$ 80 due to the fine-scale spatial sampling of the telescope aperture $\Delta P$ $\sim$ cm. However, even with a larger sampling of $\Delta P = 10$ cm, $\gamma_{ss}$ is a factor of 10. Since $N_s \sim 3 \cdot 10^4$ (per symmetric starshade quadrant), a direct FFT calculation would be computationally intractable. Similarly, the telescope zero-padding factor $\gamma_P$ per PSF term is between 8$-$16. 

\begin{figure}[h!]
  \centering
  \includegraphics[width=.9\linewidth]{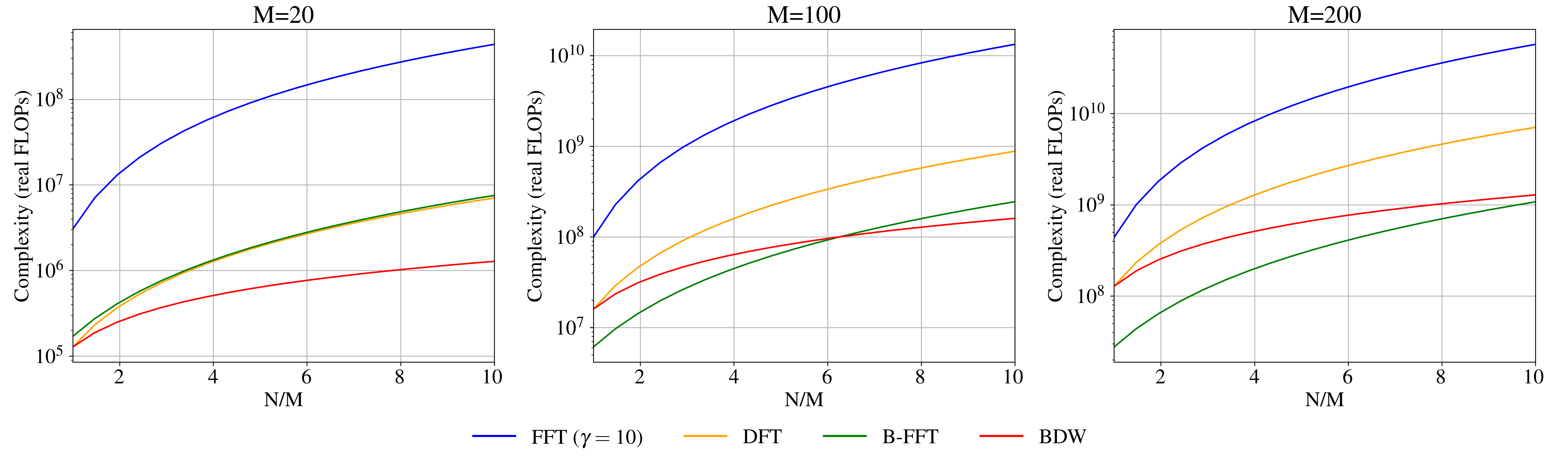}
    \caption{The relative complexity in real FLOPs among 2D Fourier transform computations described in Table \ref{tab:comparison} is shown for varying the ratio of non-zero input samples $N$ to output samples $M$ (x-axis) with complexity on a log scale (y-axis). Figures left to right show increasing the overall size of $N$ and $M$. For the (starshade $\to$ telescope)  propagation $N/M = N_s/N_P^\parallel = 20$, for the (telescope $\to$ focal plane)  $N/M = N_P/N_f =  1$. When a smaller number of output samples $M < 50$ are needed, the BDW and DFT method are optimal over the B-FFT. On the right, for larger input/output sizes $N, M \geq 200$ as used in this work, the B-FFT is generally optimal. } \label{fig: compl}
\end{figure}

In this work we use the B-FFT \citep{bailey1991fractional} to perform both Fresnel and Fraunhofer propagation. The B-FFT provides a means to compute arbitrarily spaced spectral samples of a Fourier-transform on the interval $[-\omega_{max}, \omega_{max}]$ without explicitly using zero-padding to achieve this \citep{bailey1991fractional}. The zero-padding factors $\gamma_{ss}$ and $\gamma_P$ instead act as variables in the B-FFT and can be freely varied.  The B-FFT applied for diffraction sub-sampling is described in \citet{jurling2018techniques} and in Appendix \ref{ap: bfft}. In Table \ref{tab:comparison} we summarize the computational complexity of the B-FFT approach compared to other methods, for further details and derivations see Appendix \ref{sec:complexity}. In column 2 $N$ is the number of non-zero input points, and $M$ is the number of output points. There are two stages of optical propagation. As shown in Figure \ref{fig: sampled_field}, $N$ and $M$ vary depending on what stage of optical propagation is being calculated and columns 3 and 4 represent the computational complexities evaluated for these two stages. In column 3, where the field is propagated from the starshade to telescope plane, we adopt the input-output {values} $N = N_s = 6 \cdot 10^4, M = N^\parallel_P = 3 \cdot 10^3$, we adopt $\gamma_{ss} = 40$ for the direct FFT, and $N_{quad} = 5 \cdot 10^5$ for the NU-FFT method. In column 4, where the field is propagated from the telescope aperture to the focal plane, we adopt $N  = M $ and $ N =  N_p = N_f = 250$, additionally $\gamma_P = 10$. This second stage is performed multiple times to generate the PSF basis. See Appendix \ref{sec:complexity} for a full derivation of the computational complexity.  
Figure \ref{fig: compl} illustrates the relative computational complexity among methods over a range of input-output size ratios.

\begin{table}
\centering
\caption{\textbf{Computational complexity of diffraction methods for simulating the HWO starshade.}
}
\begin{tabular}{lccccc}
\toprule
Method & Complexity & starshade $\to$ telescope [real FLOPs] & telescope $\to$ focal plane [real FLOPs]  \\
& & $(N = N_s)  \to (M = N_P^\parallel )$ &  $(N = N_P)  \to (M = N_f )$\\
\midrule
FFT & $O((\gamma \cdot N)^2 \log (\gamma \cdot N))$ &  $964 \times$ & $16 \times$ \\
\addlinespace
DFT & $O(N^2 M + M^2 N)$ & $71 \times$ & $6 \times$ \\
\addlinespace
\textbf{B-FFT} & $O((N + M)^2 \log (N + M))$ & $1 \times$ & $1 \times$ \\
\addlinespace
BDW & $O(N \cdot M^2)$ & $7 \times$ & not applicable \\
\addlinespace
NU-FFT & $O(N_{quad} + M^2 \log(M^2))$ & $ 10^{-4}\times$ & not applicable \\
\bottomrule
\end{tabular}
\tablecomments{\textbf{There are $N$ non-zero input points and $M$ output points corresponding to an optical propagation. From the starshade onwards, there are two stages of optical propagation, shown in Figure \ref{fig: starshade_im}, column 3 and column 4 show the computational complexity of each method relative to the B-FFT, for the first stage and the second stage, respectively.} These are in units of real FLOPs (floating-point operations) describing the approximate number of real adds and multiplies, assuming complex-valued inputs and outputs. }
\label{tab:comparison}
\end{table}

As shown in Table \ref{tab:comparison}, using non-uniform sampling with a NU-FFT, to compute Fresnel diffraction past a starshade mask can significantly reduce computational complexity while preserving accuracy \citep{barnett2020efficient}. The reduction in computational complexity of this approach hinges on the number of non-uniform samples needed to represent the optical field (mask and Fresnel kernel). For the starshade binary mask, the authors develop a sub-sampling scheme that utilizes the edge description of the mask. However, a scheme to generate efficient non-uniform samples of arbitrary fields, as arise in multi-stage optical pipelines, is not guaranteed to exist. Consequently, the authors do not apply the NU-FFT for the second step to compute the PSF basis. While the B-FFT is sufficient for the first step of optical propagation, it is superior to alternative methods for the second step and applicable broadly for propagating arbitrary fields.

\section{RESULTS} \label{sec: results}
The optical simulation pipeline described above is provided as an open-source tool, PyStarshade \citet{pystarshade}, described in Section \ref{sec: mission}. We present results applying the optical pipeline to perform simulations with the HWO starshade. In Section \ref{sec: solar}, we demonstrate high-fidelity imaging simulations of the solar system across viewing angles and telescope apertures. In Section \ref{sec: core_thru_results}, we further evaluate core throughput for segmented, obscured HWO apertures, showing minimal azimuthal dependence and excellent performance across wavelengths. 

\subsection{Starshade simulations} \label{sec: mission} \label{sec: starshade_mask}  \label{sec: validate}

Briefly, the PyStarshade toolbox provides implementations for optical simulations for the three starshade mission designs summarized in Table \ref{tab: mission}. The library takes a set of instrument parameters and files comprising a Design Reference Mission (DRM) and uses the \textit{propagator} class to simulate imaging.  A user-defined telescope aperture may be used, or alternatively, the propagator class can interface with HCIPy \citep{hcipy}, a coronagraph imaging library, to build a variety of pre-defined telescope apertures. 

One starshade mission concept proposes a starshade rendezvous with the Nancy Grace Roman Space Telescope (NGRST) \citep{seager2019starshade, ngrst_rendezvous} and is labeled NI2 Roman. The second starshade concept analyzed is for the Habitable Exoplanets Observatory (HabEx) \citep{hab2018}. HabEx was an earlier direct-imaging telescope mission design concept. The Habitable Worlds Observatory (HWO)  is based on HabEx \citep{hab2018} and the Large UV/Optical/IR Surveyor (LUVOIR) mission design concepts \citep{luvoir2019luvoir}. The starshade design for HWO is due to \citet{hwo_ss}. A VIS/NIR bandpass of $500- 1000$ nm has been considered for HWO, and would allow for spectral characterization of exo-Earth gases (CO2, H2O, CH4, O2, and H2) as noted above \citep{mennesson}.

\begin{table}
\centering
\caption{Optical parameters for different starshade configurations}
\begin{tabular}{lcccccc}
\toprule
Starshade Type & $\lambda$ (nm) & $d$ (m) & $\mathcal{f}$ & $N_s$ & $\theta_{P}$ (mas) & \textbf{Geometric }IWA (mas)\\
\midrule
NI2 Roman & $[425, 552]$ & $3.72 \cdot 10^7$ & $[10.7, 8.2]$ & $2.6 \cdot 10^4$ & $[45,  59]$ & 72\\
& [606, 787]& 2.61 $\cdot 10^7$ & [10.7, 8.2] &   & [65,  84] & 103 \\
& [747, 970] & 2.12 $\cdot 10^7$ & [10.7, 8.2] &   & [80, 103] & 126 \\
NW2 HabEx & [300, 1000] &  1.24 $\cdot 10^8$  & [34.8, 10.5] & $7.2\cdot 10^4$ & [19, 63] & 60 \\
HWO & [500, 1000] &  9.52 $\cdot 10^7$  & [18.9, 9.5] & $6\cdot 10^4$ & [21, 41] & 65 \\
\bottomrule
\end{tabular}
\tablecomments{NI2 ($10^{-10}$ contrast) from the SISTER toolbox \citep{sisters} with $R_{ss} = 13$ m, $D_P = 2.4$ m and 16 petals, NW2 from SISTER with $R_{ss} = 36$ m, $D_P = 4$ m and 24 petals, HWO with $R_{ss} = 30$ m, $D_P = 6$ m and 24 petals. Here the \textbf{geometric} IWA = $R_{ss}/d$. The telescope resolution is $\theta_{P} = \frac{1.2 \lambda}{D_{P}}$ and the Fresnel number $\mathcal{f} = \frac{R_{ss}^2}{\lambda d}$. The Fresnel number $\mathcal{f}$ varies with wavelength over the band and is largest at the short wavelength edge.  Here we take the focal length for HWO as $f = 10$ m.} \label{tab: mission}
\end{table}

\subsubsection{Generating a starshade mask on a grid} 
In this section, we outline the process for generating a starshade mask in the appropriate format for PyStarshade and present tests to establish the mask sampling resolution, $\Delta s$, as well as to confirm the accuracy of our diffraction simulations against NASA state-of-the art tools, diffraq (modified from fresnaq \citep{fresnaq}) and SISTER. These diffraction tests ensure that our imaging simulations match other diffraction tools as well as theoretical expectations and thereby meet the contrast required for exoplanet observations. We focus this validation on the NI2 Roman mask as other tools and analysis have focused on this example \citep{seager2019starshade, hu2021exoplanet,  barnett2020efficient, aime_ss}.

The apodization describes the fractional masking of the starshade as a function of radius that may be used to generate starshade masks with different petal numbers. Our approach requires generating a starshade mask evaluated on a uniformly spaced grid from either an apodization or a locii (set of edge points) description of a starshade. Any source of error in this step could lead to a loss of precision in the final diffraction simulations and so validating the mask accuracy is important. At a given choice of spatial sampling $\Delta s$, a binary-valued mask loses precision since the edge of the starshade will cut across pixels, therefore, we generate a grey-valued anti-aliased mask. For pixels at the edge of the starshade, the value of the pixel represents the fraction of its area that lies inside the starshade locii. Spatial anti-aliasing improves error convergence in diffraction calculations, as shown by \citet{barnett2020efficient}, and it's use here follows established methods \citep{harness2018advances}. Since the set of locii edge points must be interpolated in order to calculate the fraction of a pixel that lies within the starshade boundary, if this interpolation method is inaccurate, additional errors could be introduced in the calculated pixels. We describe our mask generation process below in more detail.

For the NI2 Roman mask, an apodization profile is provided in the file \textit{NI2.mat} from the SISTER toolbox. We use the apodization to generate a locii file, a uniformly spaced set of edge points around the edge of a starshade. The locii file is generated using the diffraq library. In particular, we generate a locii for a 16-petal mask. Note \textit{NI2.mat} further contains a locii for a 24-petal mask and the SISTER toolbox contains locii for starshades in Table \ref{tab: mission} \citep{sisters}. In order to generate the final spatially anti-aliased mask from the locii we perform the following steps: First, the discrete locii points are interpolated into a smooth, continuous boundary using a Bezier curve, implemented with the matplotlib Path module. This curve defines the starshade’s perimeter. Next, a high-resolution binary mask is constructed by sampling this curve on an upsampled grid by a factor of $4 \times 4$ relative to the final resolution. For each grid point, we determine whether it lies inside or outside the interpolated edge, assigning values of 1 (inside) or 0 (outside). Finally, spatial anti-aliasing is applied by averaging each $4 \times 4$ pixel block into a single pixel in the final mask. This averaging produces gray values at the edges of the mask. 

By varying $\Delta s$, we found that suppression stabilizes beyond $\Delta s = 1$ mm as the average of 16 pixels per gray-scale mask pixel, indicating sufficient resolution to capture the starshade’s diffraction behavior. In Figure \ref{fig: change_dx} for NI2 Roman from Table \ref{tab: mission}, the sampling is varied and the suppression profile at the telescope aperture is shown. 

To verify diffraction accuracy, we compared suppression profiles at the telescope aperture ($f^\parallel_P$) computed with our PyStarshade tool against those from the diffraq library for the NI2 starshade. Figure \ref{fig: contrast} shows differences below $10^{-10}$, confirming accuracy.
The suppression reaches down to $10^{-10}$ at this wavelength, thus demonstrating the simulation tool is sufficiently accurate to simulate imaging of systems with Earth-Sun contrasts.

\begin{figure}[h!]
  \centering
  \includegraphics[width=.6\linewidth]{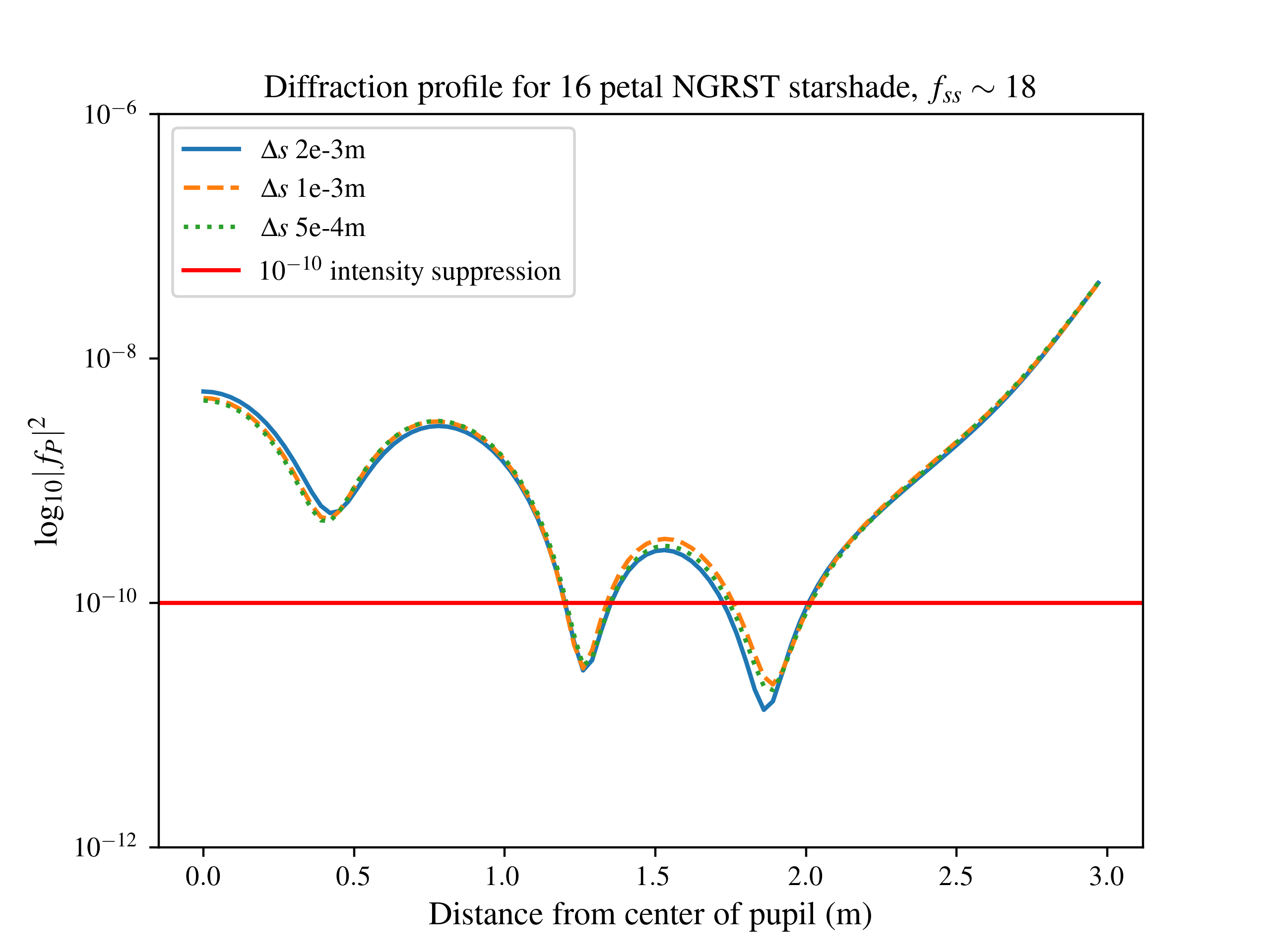}
    \caption{Suppression (diffraction intensity at the telescope aperture) of a point source is shown for the NI2 Roman starshade (16 petals) at $\mathcal{f} = 9.1$ corresponding to $\lambda = 500$ nm, with different starshade pixel sizes $\Delta s$. No change is seen increasing the discretization past $\Delta s =$ 1 mm.}
\end{figure} \label{fig: change_dx}

\begin{figure}[h!]
  \centering
  \includegraphics[width=.6\linewidth]{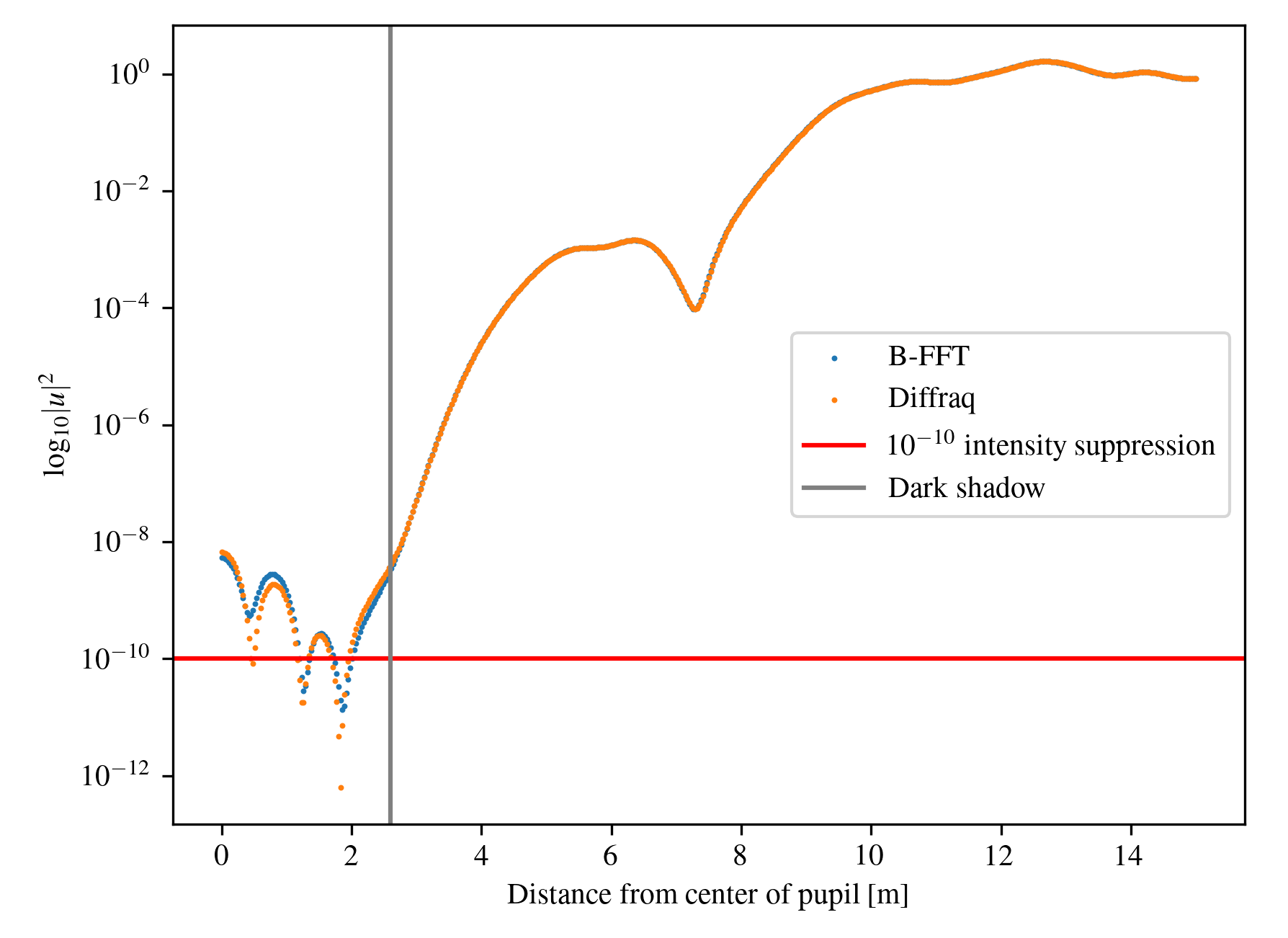}
    \caption{Suppression profile at the telescope aperture $f^\parallel_P$ is shown simulated with PyStarshade and with the diffraq library for the NI2 starshade (16 petals) at $\mathcal{f} = 9.1$ corresponding to $\lambda = 500$ nm.  The difference between these calculated profiles is below $10^{-10}$, affirming the diffraction accuracy of our simulation approach. }
\end{figure} \label{fig: contrast}

As a final validation, we test our numerical Fresnel diffraction against an analytic approximation for a circular aperture using a truncated Lommel variable expansion \citep{somm}. Adapting a script from \citet{diffraq} and following \citet{harness2018advances}, we model a gray-pixel aperture of radius $R = 30$ m. With an error tolerance of $10^{-6}$ and 50 Lommel terms as in \citet{harness2018advances}, the maximum intensity error at the telescope aperture is below this threshold, affirming the precision of our simulations.

\subsection{Solar system simulations} \label{sec: solar}
In Figure \ref{fig: ss_onaxis}, simulations of our solar system imaged with the HWO starshade described in Table \ref{tab: mission} are shown at two inclinations, face-on at 0 degrees and at 60 degrees, for both telescope apertures shown in Figure \ref{fig: hwo_pupil}. The solar system model and scenes are generated with ExoVista \citep{Stark_2022}. PyStarshade contains an example script to interface with an ExoVista scene and simulate imaging. The scenes are $1000 \times 1000$ pixels in size, with 2 mas sized pixels, covering a $2' \times 2'$ area. Simulated images are shown for a single wavelength $\lambda = 500$ nm. The PSF pixel-size $\Delta f = 2$ mas corresponds to $\sim 10 - 20$ pixels per resolution element $\frac{\lambda}{D}$. The PSF basis is simulated over the wavelength band in Table \ref{tab: mission} in steps of $\Delta \lambda = 50$ nm, for the HWO basis we have $150 \times 150$ PSF basis terms.

\begin{figure}[h!]
  \centering
  \includegraphics[width=\linewidth]{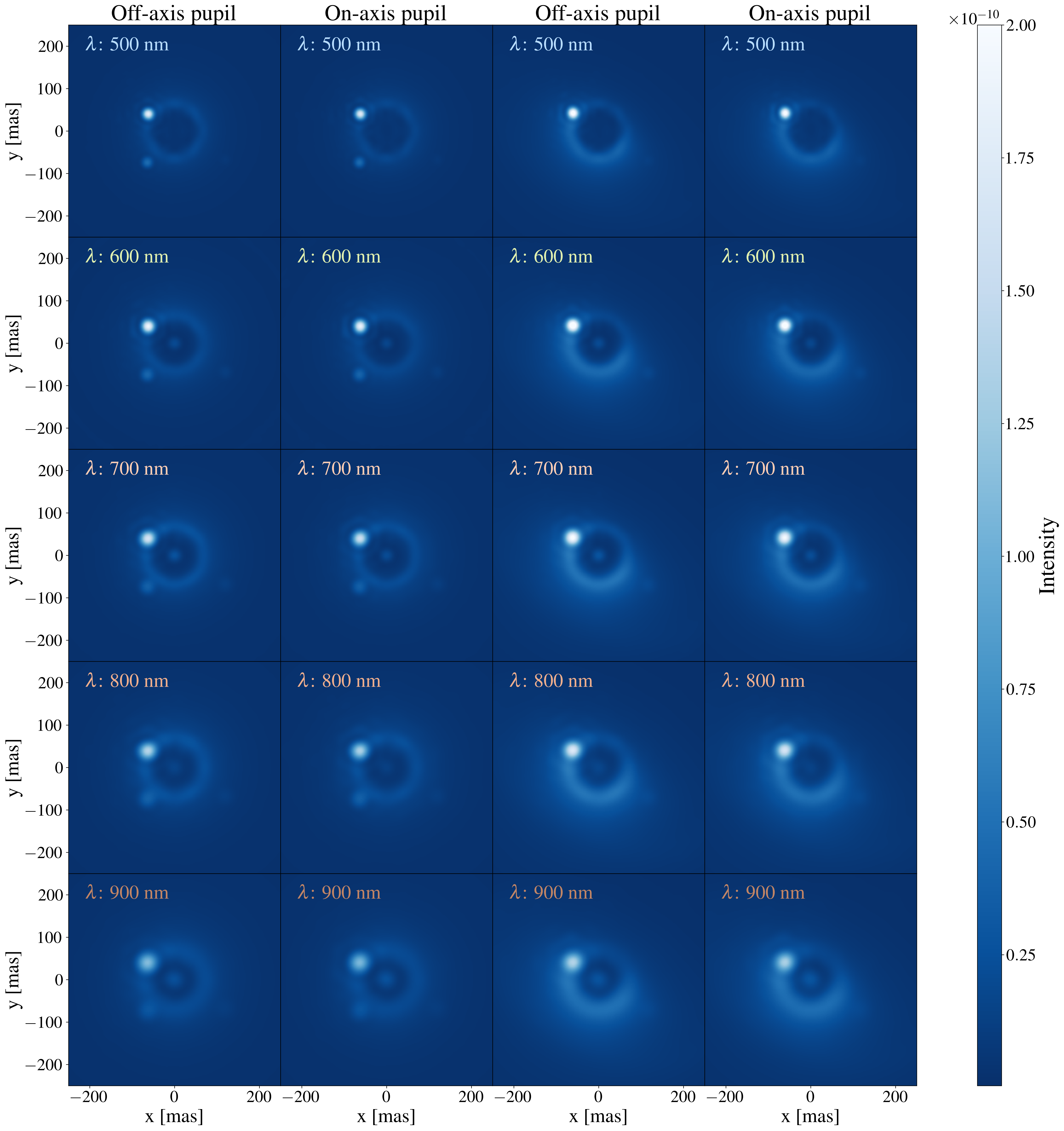}
    \caption{Solar system at 10 pc magnitude with HWO, face-on in columns 1 and 2, and at an inclination of 60 degrees in columns 3 and 4. This scene was simulated with ExoVista \citep{Stark_2022}. For each inclination, the imaging generated for each pupil type (off-axis pupil and on-axis pupil) is shown. The Fresnel number $\mathcal{f}$ changes with wavelength, hence the on-axis contrast varies slightly but is maintained below $< 10^{-9}$ over the band. Angular resolution decreases with wavelength, while the inner working angle varies minimally with wavelength.  }
\end{figure} \label{fig: ss_onaxis}
 
\subsection{Optical throughput} \label{sec: core_thru_results}
This section evaluates the optical core throughput of a starshade when paired with a segmented and obscured telescope aperture as would be used in a hybrid starshade-coronagraph mission. Generally, planet detection methods for direct-imaging utilize a matched filter with a template to locate the core of the PSF. The core throughput $c(\phi)$ for a source at position $\phi$ measures the energy in the core of the PSF as the summed intensity $I_{\phi} (x, y)$ within a photometric aperture of radius $ \frac{0.7 \lambda}{ D}$ in the focal plane, normalized by the total intensity incident on the aperture without a starshade: 
\begin{align}
c(\phi) = \frac{\sum_{\phi(x, y)\leq 0.7 \cdot \lambda / D} I_{\phi(x, y)}}{\sum_{\phi(x, y)} {I_{\phi(x,y)}^{s\mkern-7.5mu/}}}
\end{align}
The diameter $D$ that determines the photometric aperture is taken to be the the diameter of a circle with the same light collecting area as the telescope aperture, as in \citet{mennesson}. These are $D^{on-axis} = 5.25$ m and $D^{off-axis} = 5.36$ m.

\begin{figure}[h!]
  \centering
  \includegraphics[width=.7\linewidth]{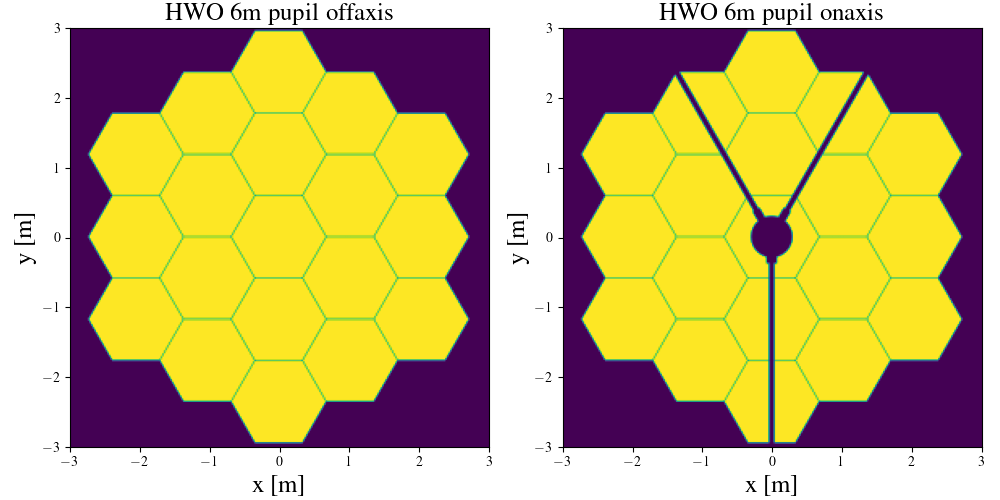}
    \caption{Baseline HWO segmented and obscured telescope apertures: 'off-axis' (left), 'on-axis' (right).  The area of the HWO on-axis aperture is 21.69, the HWO off-axis aperture is 22.53. The relative difference in area is $1 - 21.69/22.53 = 3.7\%$. } \label{fig: hwo_pupil} 
\end{figure}

A starshade achieves a theoretical 100 $\%$ throughput for off-axis sources. However, core throughput is affected by the telescope aperture. Here we compute optical throughputs for segmented entrance apertures and for both off-axis or on-axis (referring here not to the source position, but a central obscuration) telescopes to assess the complementary role of a starshade for HWO. The telescope apertures are shown in Figure \ref{fig: hwo_pupil}, provided by Rhonda Morgan (private communication 2024), a longest edge of 6 m is adopted here. 

\begin{figure}[h!]
  \centering
  \includegraphics[width=.5\linewidth]{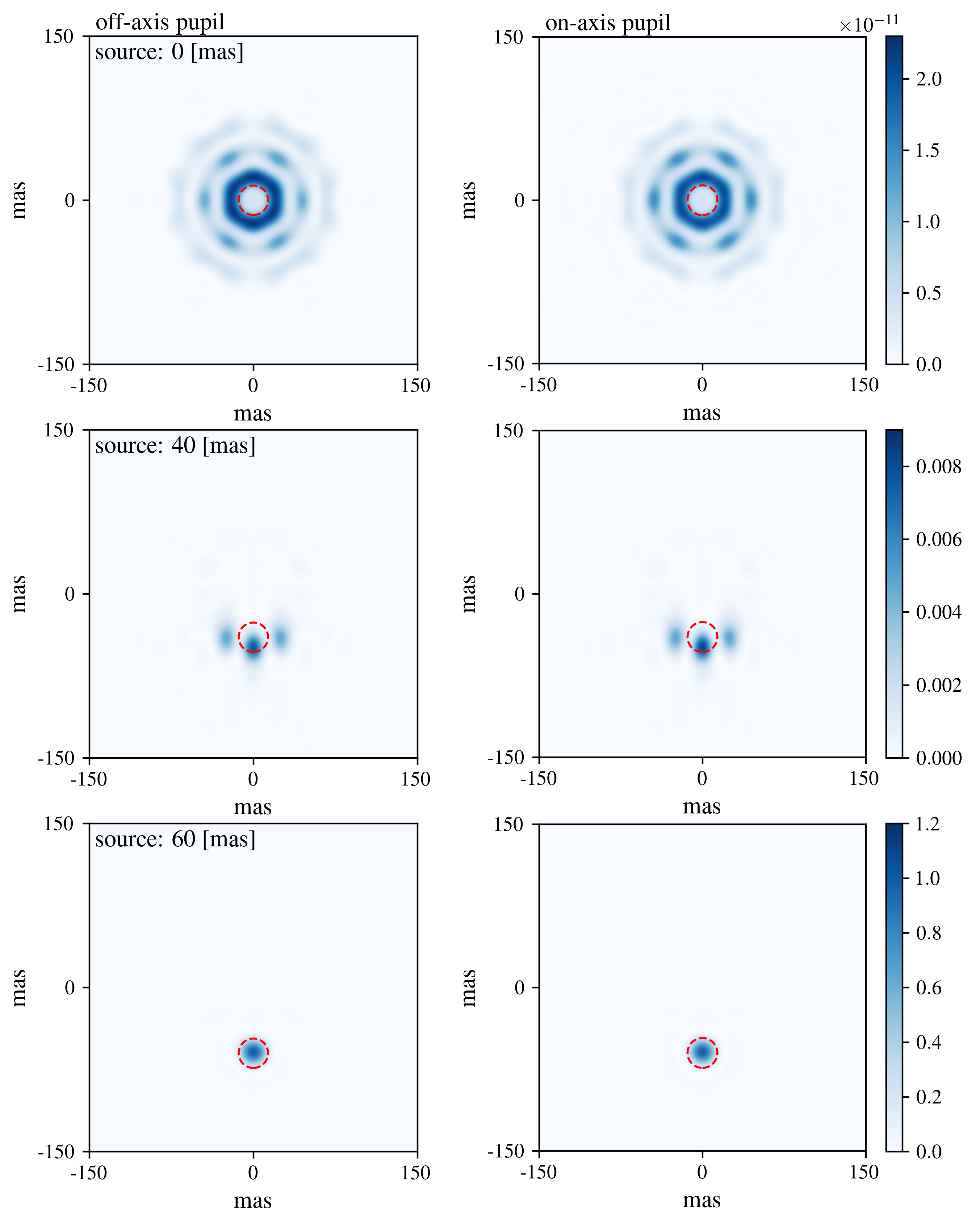}
    \caption{Example PSF's at 500 nm for the off-axis aperture (left) and on-axis aperture (right) baseline HWO apertures shown in Figure \ref{fig: hwo_pupil}. The core throughput region of radius $0.7 
    \lambda / D$ is shown by a red dashed circle, the effective diameter $D$ depends on the aperture area. Within the IWA the PSF is not centered, see \citet{aime_ss} for similar results for the 26m NGRST rendezvous starshade. In our analysis of core throughput, we center the photometric aperture on the calculated centroid of the PSF using the position that maximizes the cross-correlation with a Gaussian template of the same aperture resolution. }  \label{fig: psf}
\end{figure}

The intensity without a starshade $I^{{s\mkern-7.5mu/}}$ is calculated for each respective aperture under study. Therefore, the core throughput values capture a loss in efficiency due to the dispersion of energy in the PSF. Within the IWA the PSF is not always exactly centered, as shown in Figure \ref{fig: psf} for sources at different angular separations. Within the IWA a centroid finding method is used when estimating core throughput as described: \textbf{the} center of the PSF is found as the maximum of the cross-correlation with a Gaussian template of width given by the angular resolution of the aperture $\sigma = \lambda / D$.

\begin{figure}[h!]
  \centering
  \includegraphics[width=0.8\linewidth]{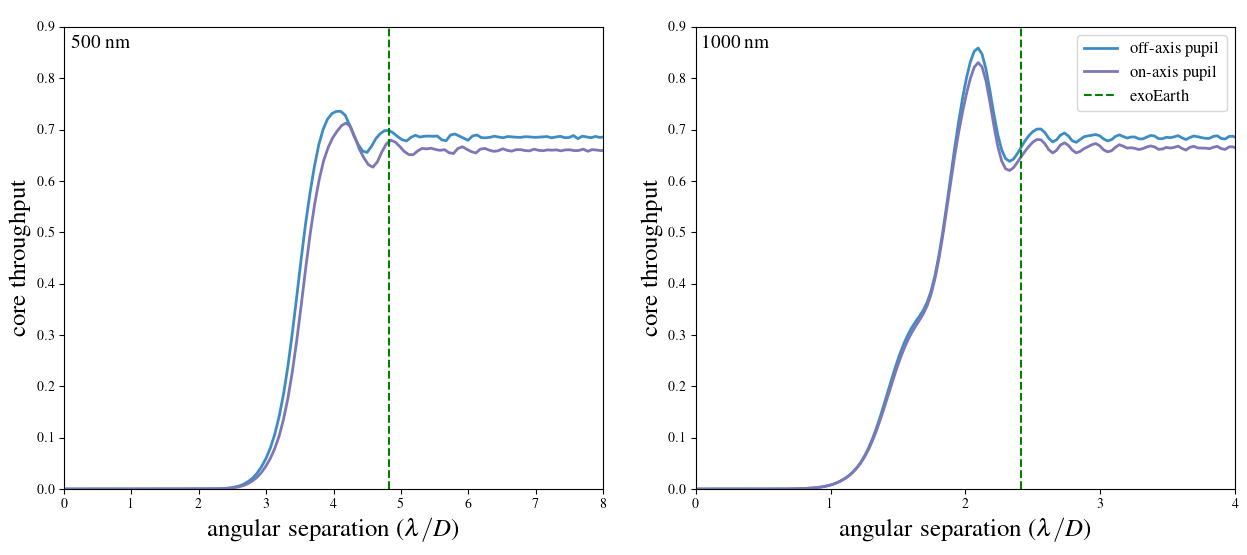}
    \caption{Core throughput within a photometric aperture of radius $0.7 \lambda / D$ at 500 nm (left) and 1000 nm (right) shown in units of angular resolution (where the effective diameters are $D^{on-axis} = 5.25$ m and $D^{off-axis} = 5.36$ m respectively). The angular separation of an exo-Earth if viewed at 12 pc, at orbital quadrature at 83 mas is shown with a green dashed line. The exo-Earth can be efficiently imaged across the entire $500-1000$ nm band.} \label{fig: throughput_iwa}
\end{figure}

\begin{figure}[h!]
  \centering
  \includegraphics[width=0.9\linewidth]{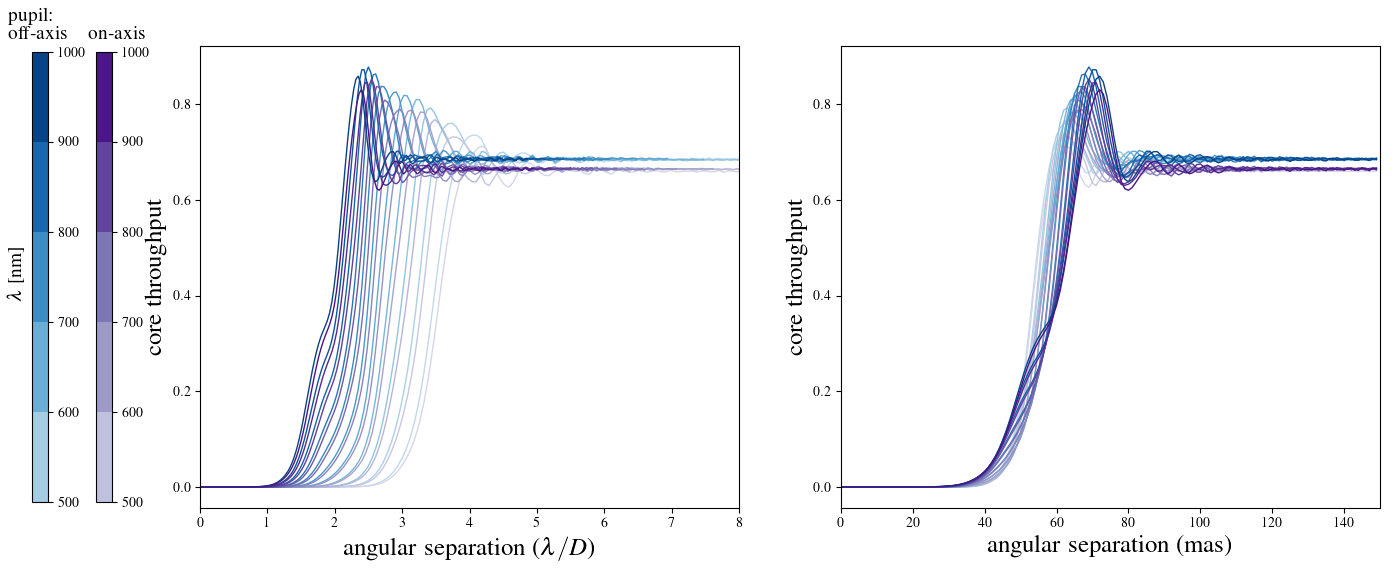}
    \caption{Core throughput within a photometric aperture of radius $0.7  \lambda / D$ for wavelengths ranging from $500$ nm to $1000$ nm in steps of $\Delta \lambda = 50$ nm, shown in units of angular resolution (left) and mas (right). These are shown overlaid for the off-axis aperture in blue and on-axis aperture in purple. Baseline HWO apertures are shown in Figure \ref{fig: hwo_pupil}. Outside of the inner working angle the core throughput is marginally lower for the on-axis aperture at $66 \%$ as compared to the off-axis aperture at $68 \%$. Inside the IWA, both apertures have the same throughput. The starshade throughput is generally not highly dependent on wavelength, however angular resolution worsens in the NIR. In the NIR, the starshade is able to image close-in planets at $2 \lambda / D$. We note the amplification of a source amplitude at the IWA, also seen in \citet{habex, hwo_ss, aime_ss}.} \label{fig: throughput}
\end{figure}

Core throughput curves with a step size of $\Delta \phi = 1$ mas, are calculated for the parameter ranges in Table \ref{tab: mission}. Shown in Figure \ref{fig: throughput} and in Figure \ref{fig: throughput_iwa} for two single wavelengths. In Table \ref{tab:core_throughput} the core throughput in the stationary region (at 130 mas) is given with varying circular thresholds. The throughput curves show no unusual deviations from expected performance and support the use of a starshade with these telescope apertures. 

In Figure \ref{fig: throughput} the core throughput is in units of mas as compared to Figure \ref{fig: throughput_iwa} where units of telescope resolution are used. The core throughput is minimally dependent on wavelength, however since angular resolution becomes broader with increasing wavelength, the IWA will be fewer units of angular resolution away from the star as wavelength increases.

\begin{figure}[h!]
  \centering
  \includegraphics[width=0.5\linewidth]{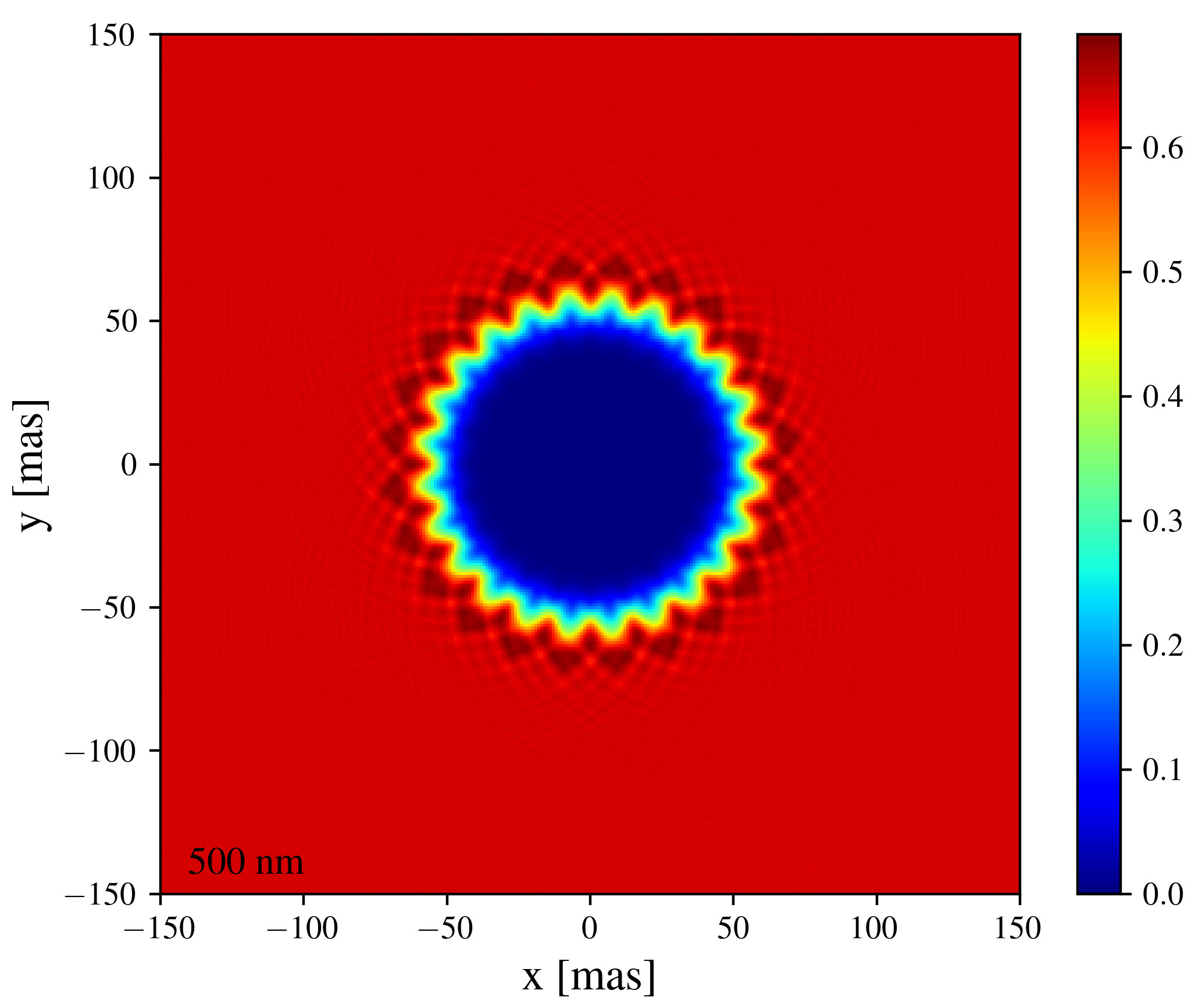}
    \caption{Core throughput in the focal plane over a grid of source positions covering $2 \cdot IWA$ at 500 nm for the on-axis aperture with support struts. Despite the asymmetric nature of the support strut, no spatial asymmetry is observed in the core throughput.} \label{fig: grid_through}
\end{figure}

To investigate any azimuthal dependence of the throughput, grids of core throughput over source positions in (x, y) were computed. An example is shown in Figure \ref{fig: grid_through}. No aperture-dependent asymmetries are observed, rather the spatial dependence reflects the diffracted field incident on the aperture.

\begin{table}[ht]
    \centering
     \caption{Core throughput outside the IWA}
    \begin{tabular}{lccc}
    \hline
    \textbf{Stationary core throughput within a radius: } \(x \cdot \lambda/D\) \\
    x:  & 0.5 & 0.7 & 1 \\
    \hline 
    HWO off-axis  & 0.47  & 0.68 & 0.82 \\
    HWO on-axis   & 0.46  & 0.66 & 0.78 \\
    \hline
    \end{tabular}
   \tablecomments{Calculated as intensity within a circle of radius $x \cdot \frac{\lambda}{D}$, where $D$ is the effective radius of each aperture, $D^{on-axis} = 5.25$ m and $D^{off-axis} = 5.36$ m. The source is at an off-axis position of 130 mas averaged over $\lambda$.  }
    \label{tab:core_throughput}
\end{table}

The IWA uses the definition in \citet{mennesson} as the angular separation at which the core throughput reaches 50\% of its maximum off-axis value. We report IWA in Table \ref{tab:IWA}.

\begin{table}[ht]
    \centering
    \caption{Calculated IWA from throughput curve.}
    \begin{tabular}{lcc}
    \hline
    \textbf{Inner working angle: } \( \) \\
    $\lambda$:  & 500 nm & 1000nm\\
    \hline   
    HWO off-axis  & 53 mas, 3.08 $\frac{\lambda}{D}$ & 57 mas, 1.68 $\frac{\lambda}{D}$\\
    HWO on-axis   & 53 mas, 3.08 $\frac{\lambda}{D}$ & 58 mas, 1.65 $\frac{\lambda}{D}$ \\
    \hline
    \end{tabular}
    \tablecomments{The IWA is reported in mas and angular resolution units over the band. Both apertures achieve close to identical IWA values. }
    \label{tab:IWA}
\end{table}
 
\section{DISCUSSION} \label{sec: discussion}

We have evaluated the B-FFT as an efficient method to perform optical imaging simulations for high-contrast imaging. The B-FFT is generally useful for diffraction simulations, and code implementing the B-FFT is provided in \citet{pystarshade}. When a zero-padding factor $\gamma > 2$ is needed to achieve the output spatial sample resolution, the B-FFT has an improved numerical complexity over the FFT. When the number of input and output samples is the same $N=M$, if $N > 50$, the numerical complexity of the B-FFT is improved over the DFT. We further describe a general optical model of a starshade in Section \ref{sec:optics}. While not singularly optimal for every stage of the optical pipeline, or for all imaging parameter ranges, we show this method to be advantageous in this setting where fine spatial features (cm scale or smaller) and asymmetries are present in the aperture. This efficiency is crucial for high-resolution simulations of next-generation telescopes like HWO, which feature complex, segmented apertures that demand precise modeling of diffraction effects.  Based on the general optical model implemented with the B-FFT, we generated a PSF basis for a 60 m HWO starshade and a 6 m segmented telescope to study core throughput performance.

Aperture radius is an important mission characteristic that can increase the yield of an exo-Earth search. Larger apertures improve both light collection and angular resolution, directly enhancing the number of detectable exo-Earths for a coronagraph. Although a 6 m aperture is part of the design baseline for HWO, 7m or even 8m apertures have also been considered \citep{stark2024pathsrobustexoplanetscience}. Thus, segmented apertures are likely to be used for larger apertures due to their stability and the ability to fold these apertures during launch \citep{seg_pupils}. Our results in Section \ref{sec: core_thru_results} show the HWO starshade used with the segmented telescope apertures achieves optimal core throughput measured within a photometric aperture of radius $0.7 \lambda / D$ (where $D$ is taken to be an equivalent diameter corresponding to the area of the telescope aperture). In particular, the off-axis, segmented aperture and on-axis obscured and segmented aperture achieve a $68 \%$ and $66 \%$ core throughput, respectively, compared to an ideal circular aperture with a $ 68 \%$ core throughput.

This core throughput measure is used as a benchmark for standardized comparisons of starlight suppression systems \citep{mennesson} as it approximately maximizes the SNR against an unknown background for a circular aperture at expected noise levels \citep{Stark_2014}.
Core throughput captures both the redistribution of light outside of the central core of the PSF due to the aperture shape, and the diffraction effects of the starshade. The reduction for the on-axis aperture is attributable to the central obscuration, not the diffraction features of the starshade. The exposure time is approximately inversely proportional to the core throughput. In \citep{mennesson} for spectroscopic observations of an Earth-like exoplanet at 12 parsecs with a circular 6-meter telescope, a starshade requires 1-2 weeks of exposure time. For an on-axis aperture this corresponds to a minimal increase in exposure time of $3 \%$.

As described in \citet{vvc}, the core throughput measure may be overly conservative, as one may be able to make use of the side lobes of the PSF. This may be especially true for a starshade, where the dark shadow is created upstream of the aperture and the instrument does not suffer from speckle error, meaning a forward model of the instrument may be realistic.  Starshade specific data processing is explored in \citet{data_challenge}, focused on background estimation using principal component analysis and independent component analysis. High contrast images are photon-limited and thereby subject to data-dependent photon noise. \cite{hu2021exoplanet} derive a detector for the photon-limited Poisson regime which utilizes a model PSF basis and jointly estimates the unknown background. \citet{ahmed2023exoplanet} use a neural network for detection, trained on data from SISTER \citep{sisters}. Both simulated data and the detection data processing strategies in these works do not model segmented apertures. Future work includes using the HWO PSF basis to assess whether model-based matched filtering can utilize the full PSF.  

The inner working angle (IWA) provides a concise metric for comparing the projected star-planet separation at which the planet can be observed with different starlight suppression systems and is approximately proportional to the starshade radius $R_{ss}$ and lateral flight distance $d$ as $IWA = \frac{R_{ss}}{d}$ \citep{iwa_cady}. With these faster simulation techniques, a more precise wavelength and aperture-dependent IWA can be obtained from core throughput. We determine the IWA is the same for both apertures considered, the IWA varies by less than 5 mas over the  500-1000 nm band, however provides a tighter imaging angle from $53$ mas to $58$ mas as compared to the coarse IWA of $65$ mas quoted for the HWO starshade. Angular resolution changes with wavelength, as $\frac{\lambda}{D_P}$ and we find the IWA is approximately $1.7 \frac{\lambda}{D_P}$ resolution elements at the NIR and $3 \frac{\lambda}{D_P}$ at the visible. Although our reported IWAs differ by only a few milliarcseconds, such variations translate into a large gain in accessible phase space and since planets are spend most of their orbit at smaller projected separations than quadrature, this can materially affect observability. Furthermore, as described in \citet{Stark_2020}, the volume of candidate direct-imaging targets scales proportionally to IWA$^{-3}$. Hence, a small improvement in IWA also increases the number of potential targets. However, dimmer targets further away may require longer exposure times.  

Tools like SISTER provide a broader simulation framework, including noise sources like solar glint and local zodiacal light. PyStarshade is flexible and focuses on high-resolution diffraction simulations.  We leave it to future work to include a full detector model, as well as sources of noise. The brightness from local zodiacal light will add a spatially smooth component to imaging with a median value of  23 $\mathrm{mag\,arcsec^{-2}}$ at visible wavelengths \citep{mag}. Future work will use the core throughput metrics presented here to calculate precise exposure times for an Earth-Sun analog. Additionally, exploring model-based matched filtering approaches tailored for HWO apertures may help mitigate throughput losses. The B-FFT framework and PyStarshade tools developed here provide a foundation for comprehensive studies of diffraction-limited high-contrast imaging in next-generation space telescopes.

\section{CONCLUSIONS} \label{sec: conclusion}
In this work, we introduced the B-FFT for computationally efficient direct-imaging simulations of exoplanets involving multiple stages of optical propagation. This method and general diffraction tools are implemented in the $PyStarshade$ package \citep{pystarshade}. We describe a general optical diffraction model of a starshade and the associated numerical challenges due to fine-scale features of a starshade and telescope aperture. 

We use $PyStarshade$ to simulate imaging of a 60 m starshade flown with a 6 m HWO telescope. In particular, we simulate the effect of segmented and obscured apertures on imaging and simulate a PSF basis. We report core throughput curves between 500-1000 nm for segmented and obscured HWO baseline telescope apertures. We find a core throughput of $68\%$ for a segmented off-axis telescope and $66\%$ for a segmented on-axis telescope. This suggests a starshade achieves close to (or at) the optimal core throughput of a circular aperture, and in terms of this performance metric is highly complementary to HWO for exoplanet characterization. We also evaluated inner working angles (IWA), showing that the IWA remains consistent across segmented and unobscured apertures, varying from $3.1 \frac{\lambda}{D}$ at 500 nm to $1.7 \frac{\lambda}{D}$ at 1000 nm. 
Finally, we demonstrate the full-fledged image simulation capabilities with simulations of our solar system over the HWO band. Simulations of this form are critical towards developing image post-processing tools, evaluating exoplanet retrieval efficiency and thereby the complementary role of a starshade for HWO. 

This approach shows reduced computational complexity over the FFT or DFT in the context of starshade simulations. In particular, when a zero-padding factor $\gamma > 2$ is needed to achieve the output spatial sample resolution, the B-FFT has an improved numerical complexity over the FFT. When the number of input and output samples is the same $N=M$, if $N > 50$, the numerical complexity of the B-FFT is improved over the DFT, reducing the complexity of high-fidelity optical simulations.

\acknowledgments
We thank Stuart Shaklan for his insightful feedback on this work.

\software{Fresnaq \citep{fresnaq}, diffraq \citep{diffraq}, ExoVista \citep{Stark_2022}, SISTER \citep{sisters}}

\appendix

\section{Bluestein FFT} \label{ap: bfft}
The Bluestein FFT \citep{bluestein, bailey1991fractional} (related to the Chirp-Z transform \citep{chirpz}) is a computationally efficient method to compute DFT samples of a compact input signal over a chosen interval of uniformly spaced output frequencies and is used here to calculate Fresnel and Fraunhofer diffraction. We briefly describe this method applied here for 2D FFT's, or a more expansive description see \cite{jurling2018techniques}. In Section \ref{sec:optics} we described the continuous optical model, here we describe the discrete implementation of this model. We switch from continuous coordinates to discrete input spatial indices $x, y \to n_x, n_y \in\{ {0, 1, \ldots \gamma \cdot N-1} \}^2$ and output spatial indices $\omega_x, \omega_y \to k_x, k_y  \in\{ {0, 1, \ldots M-1} \}^2$. We consider the input signal $f[n_x, n_y]\in \mathbb{C}^{\gamma \cdot N \times \gamma \cdot N}$ to be nonzero over a region smaller than $N\times N$ spatial samples, however zero-padded to size $\gamma \cdot N \times \gamma \cdot N$. The 2D DFT of $f[n_x, n_y]$ is defined as:

\begin{equation} 
F[k_x, k_y] = \sum_{n_x=0}^{\gamma \cdot N-1} \sum_{n_y=0}^{\gamma \cdot N-1} f[n_x, n_y] e^{-j 2 \pi \frac{k_x n_x + k_y n_y}{\gamma \cdot N}},
\end{equation} \label{eq: 2ddft}
In Equation \ref{eq: 2ddft}, the indices \(k_x, k_y\) correspond to frequency samples \(\left( \frac{k_x}{\gamma \cdot N\Delta x}, \frac{k_y}{\gamma \cdot N \Delta y} \right)\) where $\Delta x$ and $\Delta y$ are the input spatial sampling. 

Using the identity \(2 (n_x k_x + n_y k_y) = (n_x^2 + n_y^2) + (k_x^2 + k_y^2) - (n_x - k_x)^2 - (n_y - k_y)^2\) and reducing the sum to the non-zero terms the DFT can be rewritten as:
\begin{equation}
F[k_x, k_y] = e^{-j \pi \frac{k_x^2 + k_y^2}{\gamma \cdot N}} \sum_{n_x=0}^{N-1} \sum_{n_y=0}^{N-1} f[n_x, n_y] e^{-j \pi \frac{n_x^2 + n_y^2}{\gamma \cdot N}} e^{j \pi \frac{(n_x - k_x)^2 + (n_y - k_y)^2}{\gamma \cdot N}}.
\end{equation}
Define the sequences:
\begin{equation}
a[n_x, n_y] = e^{-j \pi \frac{n_x^2 + n_y^2}{\gamma \cdot N}}, \quad b[n_x, n_y] = e^{j \pi \frac{n_x^2 + n_y^2}{\gamma \cdot N}}, \quad c[n_x, n_y] = f[n_x, n_y] a[n_x, n_y]
\end{equation}
$n_x, n_y \in \{ 0, 1, \ldots, N-1 \}$
The DFT can equivalently be expressed as:
\begin{equation}
F[k_x, k_y] = a[k_x, k_y] \cdot \left(c ** b \right)[k_x, k_y]
\end{equation}
where \((c ** b)[k_x, k_y]\) denotes the 2D convolution of \(c[n_x, n_y]\) and \(b[n_x, n_y]\) evaluated at samples \([k_x, k_y]\). 

Next, we note that since we require $M$ output samples and our signal support is length $N$, the sequences are zero-padded to size $a, b, c \in \mathbb{C}^{(N + M - 1) \times (N + M - 1)}$ in order to calculate a circular convolution. The convolution is computed with 2D FFT's (where here we denote an FFT with $F_{FFT}$):
\begin{align}
(c**b)[k_x, k_y] = F_{FFT}^{-1} [F_{FFT}[c] \cdot F_{FFT}[b]][k_x, k_y]
\end{align}
Although this produces $(N+M-1) \cdot (N+M-1)$ values, only $M\cdot M$ correctly correspond to the desired DFT samples. $F_{FFT}(b)$ is analytically known so does not need to be computed. 

The output frequency spacing of the B-FFT is determined by the choice of $\gamma$ in the definitions of the chirp sequences $a, b, c$, not by the size $N$ of the input $f$,  or the size $M$ of the output $F$.  Hence $\gamma$ may be somewhat arbitrarily modified without affecting the computational complexity. The sampled output frequencies need not be centered (i.e., \(k_x, k_y\) can be offset from zero). Here, the effective zero-padded length value \(\gamma \cdot N \) is chosen based on the desired spatial sampling of the output as described in Section \ref{sec: pupil_sampling}.

\section{Out-of-core Bluestein FFT} \label{ap: oom}
In optical applications, it is quite common to need to perform large 2D FFTs that exceed available memory. We present a method to serially chunk this 2D FFT into smaller FFTs that fit in memory, allowing the desired FFT to be computed cumulatively. This technique utilizes the B-FFT applied to chunked inputs and is implemented in $PyStarshade$.

Chunking an FFT over partial regions of the input requires full-size zero-padding per partial FFT to resolve the same output frequencies and therefore this approach is not viable for parallelizing an FFT. 
However, using a B-FFT one only needs to zero-pad each input chunk by the desired number of frequency samples $N_P$. This has a utility in the context of a starshade (where $N_s$ can be large) and computing an FFT of size $(N_s + N_P) \times (N_s + N_P)$ may quickly hit a memory bandwidth. 

For a chunking factor of $c$ (the divisor of the input), we compute $c$ B-FFT's each with an input size of $N_P + \frac{N_s}{c}$ and complexity $O(2\cdot (N_P + \frac{N_s}{\sqrt{c}})^2 \log (\frac{N_s}{c} + N_P))$. Therefore the overall complexity is $O(2 \cdot (\sqrt{c}N_P + N_s)^2 \log (\frac{N_s}{\sqrt{c}} + N_P))$. There is a small overhead as compared to computing a single Bluestein FFT, however, the chunked approach allows for full memory utilization.

This chunking approach generally exhibits slower performance compared to performing a full-scale out-of-core Bluestein FFT (B-FFT), where the 2D FFT computation is directly decomposed into sequential 1D FFT stages: first computing the 1D FFTs of all rows, then performing the 1D FFTs of all columns, with intermediate data transfers to and from memory storage. However, the advantage of chunking lies in avoiding the synchronization overhead associated with completing the FFT computations across all rows before proceeding. Furthermore, each chunked B-FFT computation directly corresponds to the physical optical propagation of a specific segment of the starshade, whereas individual 1D FFT computations (over rows or columns) in the full-scale decomposition do not have a direct physical interpretation. In this way, a specific segment of the starshade can be modified and propagated alone without needing to propagate the full-sized mask.

Lastly, it's worth noting that a DFT, as a matrix multiplication, inherently supports chunking without any increase in complexity.

\section{Computational complexity} \label{sec:complexity}

We use the sampling relations and values described in Section \ref{sec: pupil_sampling} to evaluate computational complexities for i) Fresnel propagation from starshade to telescope aperture to compute $f_P^\parallel$, and ii) simulating a PSF basis $\{\psi_{\lambda, k} \; : \; k \in K \}$. Here we consider the complexity in the context of HWO starshade simulations described in detail below in Section \ref{sec: mission} with instrument parameters in Table \ref{tab: mission} and compare our approach to other standard diffraction methods. Both steps (i) and (ii) are of comparable computational complexity in the optical pipeline and results are summarized in Table \ref{tab:comparison}. We find the B-FFT to be efficient for both these calculations individually. 

The geometric dark shadow is approximately the same diameter as a starshade, with $\Delta P = 2$ cm, we require $N_P^\parallel \sim 3000$. 
We adopt $N_P = 250$, $N_s = 60,000$, $IWA \sim 55$ mas, each PSF is of size $N_f = 250$ pixels along one axis. We assume zero-padding factors of $\gamma_{ss} =  40$, and $\gamma_P =  10$. The total number of PSF basis terms is $K = 10^4$. An ideal starshade is symmetric and this property reduces the computational complexity in practice. For the calculations here, symmetry not change the relative computational complexity. For simplicity and generality for non-symmetric masks or perturbations, symmetric reductions are not applied in the complexity analysis. 

We briefly summarize the complexity of a B-FFT used for imaging, for $N$  non-zero input samples (over the starshade mask or telescope aperture) and $M$ output samples (aperture field or focal plane). The B-FFT is calculated with two 2D FFT's, each of size $(N+M)\times(N+M)$. Since the time-complexity of a 2D FFT of size $N \times N$ is $O(N^2 \log N)$ \citep{golub1983matrix}, the overall complexity of the B-FFT is $O((N + M)^2 \log (N + M))$ \citep{bluestein, bailey1991fractional}.

Once the pupil field $f_{P}^\parallel(\mathbf{x})$ has been computed for a starshade, it can be used to generate PSFs for various telescope apertures. Using the B-FFT to calculate $f_P^\parallel$ in (i) is of complexity $O((N_s + N_P^\parallel)^2 \log (N_s + N_P^\parallel)$ and dominated by the number of non-zero points inside the starshade ($N_s = 10 N_P$). The complexity of directly using a 2D FFT for this calculation is $O((\gamma_{ss} \cdot N_s)^2 \log \gamma_{ss} N_s)$ \citep{goodman2005introduction}, where $\gamma_{ss}$ is the zero-padding factor of $N_s$ starshade samples along one axis. Alternatively, as in \cite{hu2017simulation}, without the use of an FFT, one may calculate the pupil field using a DFT matrix multiplication (DFT) of complexity $O(N_s^2 N_P^{\parallel} + {N_P^{\parallel}}^2 N_s)$. Both the direct FFT and DFT are not computationally optimal in this setting. The boundary-diffraction wave (BDWF) method \citep{cady2012boundary} computes an edge integral over the starshade and is of approximate complexity $O(N_s \cdot {N^{\parallel}_P}^2)$ per output point in the pupil plane and approximately $2N_s$ points around the starshade mask are summed. Based on stated complexities, if a coarser aperture sampling is used with $\Delta P > 8$ cm and $N^\parallel_P < 1600$, BDWF is advantageous over the B-FFT. However, for the sampling used here the B-FFT is comparable. The complexity of quadrature and the NU-FFT for the 26 m NGRST NI2 starshade mask diffraction is reported as $O(N_{quad} + {N^{\parallel}_P}^2 \log({N^{\parallel}_P}^2))$  \citep{barnett2020efficient}, where $N_{quad} \sim 5\cdot 10^5 $ is the number of quadrature points over the starshade (comparable to the number of points along one axis of the starshade with cartesian sampling).  This approach may be used to compute $f^\parallel_P$, which can then be loaded into PyStarshade to simulate a PSF basis for a chosen aperture. 

Second (ii) we consider simulating the PSF basis $\{ \psi_{\lambda, k}(\mathbf{x})\}$. Here we assume the same number of input and output points $N = M$ as $N_P = N_f$. The DFT has optimal complexity when $N < 44$.

In Table \ref{tab:comparison} we report the comparison applied to optical propagation steps in real floating point operations. These are calculated as FFT: $10(\gamma \cdot N)^2 \log (\gamma \cdot N)$, DFT: $8(N^2 M + M^2 N)$, B-FFT: $20(N + M)^2 \log (N + M)$ and BDW: $16 N \cdot M^2$ \citep{golub1983matrix}.

Since the requisite contrast levels are of order $10^{-10}$, in optical propagation calculations 64-bit precision (per real, and imaginary value) is used here. Computing FFT's of a large starshade mask at the requisite precision may not fit in RAM, B-FFT's provide a way to chunk the propagation calculations with a small overhead on complexity. This is described in Appendix \ref{ap: oom}.
The imaging outside of the $2\cdot IWA$ region requires a linear convolution with a single off-axis PSF. This is efficiently performed for all pixels in the source field with a two-step FFT, where both source field and PSF are zero-padded to twice the size of the source field for correctness. This calculation makes a negligible contribution to the overall computational cost.


\bibliography{sample63}{}
\bibliographystyle{aasjournal}

\end{document}